\documentclass[useAMS,usenatbib]{mn2e}

\newcommand{\ltsimeq}{\raisebox{-0.6ex}{$\,\stackrel 
        {\raisebox{-.2ex}{$\textstyle <$}}{\sim}\,$}}

\begin{document}

\title[The 3-D clustering of radio galaxies in the TONS survey]
%{The 3-D clustering of radio galaxies in the TONS survey: redshift spikes and their implications for radio galaxy bias}
{The 3-D clustering of radio galaxies in the TONS survey}

\author[Brand et al.]{Kate Brand$^{1,2\star}$,Steve Rawlings$^{2}$,Gary J. Hill$^{3}$, Joseph R. Tufts$^{3}$\\
$^1$National Optical Astronomy Observatory, Tucson, AZ 85719-6732, USA \\
$^2$Astrophysics, Department of Physics, Keble Road, Oxford OX1 3RH, UK \\
$^3$McDonald Observatory and Department of Astronomy, University of Texas at Austin, RLM 15.308, Austin, TX 78712, USA\\
%$^4$SIRTF Science Center; MS-220-6; California Institute of Technology, 1200 E. California Boulevard, Pasadena, CA 91125\\
}
\maketitle

\begin{abstract}
\noindent
We present a clustering analysis of the Texas-Oxford NVSS Structure (TONS) radio galaxy redshift survey. This complete flux-limited survey consists of 268 radio galaxies with spectroscopic redshifts in three separate regions of the sky covering a total of 165 deg$^2$. By going to faint radio flux densities ($s_{\rm{1.4}}\ge$3 mJy) but imposing relatively bright optical limits ($E\approx R\approx$19.5), the TONS sample is optimised for looking at the clustering properties of low luminosity radio galaxies in a region of moderate (0 $\ltsimeq$ $z$ $\ltsimeq$ 0.5) redshifts. We use the two point correlation function to determine the clustering strength of the combined TONS08 and TONS12 sub-samples and find a clustering strength of $r_0(z)$=8.7$\pm$1.6 Mpc ($h$=0.7). If we assume growth of structure by linear theory and that the median redshift is 0.3, this corresponds to $r_0(0)$=11.0$\pm$2.0 Mpc which is consistent with the clustering strength of the underlying host galaxies ($\sim 2.5 L\star$ ellipticals) of the TONS radio galaxy population.
\end{abstract}


\footnotetext{$^{\star}$Email: brand@noao.edu}

\section{INTRODUCTION}
\label{sec:intro}

Radio galaxies are ideal probes of large-scale structure as they are biased tracers of the underlying mass and can be easily detected out to high redshifts. By using biased galaxies populations, one can efficiently trace huge super-structures (i.e clusters of clusters of galaxies) which are still in the linear regime and can therefore be directly traced back to rare fluctuations in the initial density field at recombination. However, in order to be useful probes, it is vital to understand how different populations of radio galaxies trace the underlying dark matter (i.e. their bias) and how this has evolved with time. 

Different populations of galaxies demonstrate a range of clustering strengths as measured by the clustering length, $r_0$. For example $r_0$ $\approx$ 8.1$\pm$0.9 Mpc ($h$=0.7) for $L_\star$ elliptical galaxies \citep{nor02} and $r_0$ $\approx$ 5.4 Mpc for IRAS selected galaxies \citep{fis}. The power-law fit to the correlation function can also be applied to galaxy cluster surveys. For clusters of galaxies selected from the APM survey, the correlation length has been measured to be $r_0$ $\approx$ 20.4$\pm$3.4 Mpc for Abell Richness Class (ARC) $>$1 clusters \citep{dal} and $r_0$ $\approx$ 30.4 Mpc for ARC $>$2 \citep{crof}. Hence as one would expect from high-peak bias effects \citep{kai}, clusters are more strongly clustered than galaxies. The clustering of radio galaxies will follow that of the underlying host galaxy population but it may also be affected by the radio triggering mechanism in different populations and/or environments. 

%add in relation of bias with lstar?

Because of the difficulties in obtaining large, complete radio galaxy redshift surveys, the clustering properties of radio galaxies are relatively poorly constrained. For local radio galaxies, \citet{pn} find a correlation length of $r_0$ $\approx$ 15.7 Mpc ($h$=0.7). \citet{mag04} calculate $r_0$ = 13.0$\pm$0.9 Mpc for 2dFGRS/FIRST radio galaxies with AGN signatures in their optical spectra (before correcting for redshift space distortions). These radio galaxies therefore cluster with a strength between that of normal galaxies and rich clusters of galaxies. This is consistent with their preferential location in poor groups of galaxies. By deprojecting the angular correlation function of NVSS radio sources, \citet{bw} find a correlation length of $r_0$ $\approx$ 8.6 Mpc. In this case, the clustering signal comes from radio galaxies with a median redshift of $z\sim$1 but the correlation function has been corrected for growth of clustering under linear theory \citep{pee}. %To determine any evolution in the clustering and bias of radio galaxies with redshift, one must be careful that the comparison samples are comparable in their selection criteria and radio luminosity. In particular, previous radio galaxy redshift surveys have typically been conducted with flux density limits higher than that adopted for the survey presented in this paper (\citealt{pn}; \citealt{lac}; \citealt{benn}; \citealt{hr}). 

With the advent of large area radio surveys down to fainter flux densities (\citealt{con}; \citealt{bwh}), we can define lower radio flux density limited surveys over large areas of the sky. We have defined a new radio galaxy redshift survey: the Texas-Oxford NVSS Structure (TONS) survey. This survey comprises the largest complete spectroscopic sample of low luminosity (predominantly FRI) radio galaxies over large contiguous areas yet compiled.  By going to fainter radio flux densities ($s_{\rm{1.4}}\ge$3 mJy) and imposing optical limits ($E\approx R\ltsimeq$19.5), the TONS survey is optimised for looking at clustering of objects in a region of moderate (0 $\ltsimeq$ $z$ $\ltsimeq$ 0.5) redshifts. The radio flux density limit of $s_{\rm{1.4}}\ge$3 mJy increases the space density of radio galaxies powered by AGN but does not go so deep that the sample becomes dominated by galaxies whose radio emission originates from star formation mechanisms.

In this paper, we determine the clustering strength of the low luminosity TONS radio galaxies at moderate redshifts. From the correlation length we will be able to estimate the bias of this population and compare it to previous estimates of both higher luminosity radio galaxies and similar radio galaxies in the local Universe. By cross-matching radio surveys with large spectroscopic surveys such as the Sloan Digital Sky Survey (SDSS; \citealt{aba04}) and 2dF galaxy redshift survey \citep{col} with NVSS and FIRST, even larger samples can now be compiled (see e.g., Jarvis, Clewley \& Brand in prep.). 

The paper will be structured in the following way: In Sec. 2 we summarise the selection techniques used in the Texas-Oxford NVSS Structure (TONS) redshift survey of radio galaxies: this technique has already been discussed in detail for the TONS08 sub-sample \citep{brand}. We present the full TONS12, TONS08w and TONS16w sub-samples and their spectra in Sec. 3. Sec. 4 summarises the method used to obtain a model redshift distribution and compares this to the observed redshift distribution of each of the TONS sub-samples. In Sec. 5 and Sec. 6, we describe the method used to calculate the two point correlation function and present the results for the different sub-samples. Sec. 7 is a discussion which compares the results to previous work. The main conclusions are summarized in Sec. 8.  

Unless otherwise stated, we assume a spatially flat $\Lambda$CDM Universe throughout the paper with the following values for the cosmological parameters:
Hubble constant: $H_{0}=70~ {\rm km~s^{-1}Mpc^{-1}}$; $h=H_{0}/100=0.7$; matter density parameter at $z$=0: $\Omega_ {\rm M}(0)=0.3$; vacuum density parameter at z=0: $\Omega_ {\Lambda}(0)=0.7$. 

\section{SAMPLE SELECTION}
\label{sec:selection}

The Texas-Oxford NVSS Structure (TONS) survey comprises three independent regions on the sky selected in the same areas as the 7CRS \citep{wil02} and the TexOx-1000 (TOOT) survey \citep{hr}. The TONS08 (Texas-Oxford NVSS Structure 08$^h$) sub-sample is the subject of \citet{brand}. The TONS12, TONS16w and TONS08w surveys are presented in this paper. The sky regions surveyed by these sub-samples are shown in Table.~\ref{tab:radec}. Table.~\ref{tab:samples} summarises the properties of the different TONS sub-samples in addition to the various radio galaxy samples that are used or referred to throughout this paper. TONS08 and TONS12 have the faintest radio flux densities limit ($s_{\rm{1.4}}\ge$3 mJy), while TONS08w and TONS16w were designed to survey larger areas but due to limited telescope time, were less radio deep ($s_{\rm{1.4}}\ge$30 mJy). TONS08w was specifically designed to determine the spatial extent of the TONS08 super-structures which appeared to be unbounded by the TONS08 survey \citep{brand}.

\begin{table*}
\begin{center}
\begin{tabular}{rrr}
\hline\hline
sample & RA & Dec\\
\hline
TONS08                    & $08^h10^m20^s\le$ RA $\le08^h29^m20^s$ & $24^\circ10^\prime00^{\prime\prime}\le$ DEC $\le29^\circ30^\prime00^{\prime\prime}$\\
TONS12                    & $12^h48^m00^s\le$ RA $\le13^h12^m00^s$ & $33^\circ00^\prime00^{\prime\prime}\le$ DEC $\le38^\circ00^\prime00^{\prime\prime}$\\
TONS08w                   & $08^h00^m00^s\le$ RA $\le08^h41^m00^s$ & $21^\circ50^\prime00^{\prime\prime}\le$ DEC $\le31^\circ50^\prime00^{\prime\prime}$\\
TONS16w                   & $16^h24^m00^s\le$ RA $\le16^h56^m00^s$ & $42^\circ00^\prime00^{\prime\prime}\le$ DEC $\le51^\circ00^\prime00^{\prime\prime}$\\
\hline\hline
\end{tabular}
{\caption[]{\label{tab:radec} The RA and DEC limits for the different TONS sub-samples.}}
 \end{center}
 \end{table*}

Unlike the low-frequency selected 7CRS and TOOT, the TONS survey is selected at 1.4 GHz from the NVSS. For objects of typical spectral index, the TONS08 and TONS12 sub-samples go to fainter radio flux densities than TOOT (which has a 151 MHz flux density $s_{\rm{151}}$ limit of 100 mJy, corresponding to $s_{\rm{1.4}}\approx$20 mJy for radio spectral index\footnote{$\alpha$ is the spectral index for radio sources where the radio flux density $s_{\rm{\nu}} \propto \nu^{-\alpha}$, where $\nu$ is the observing frequency}$\alpha\approx$0.8). In addition, TONS has an optical magnitude limit ($E\approx R\approx$19.5) and colour cut ($B$-$R >$1.8) imposed on it. This means that we can efficiently trace large-scale structure at moderate ($z<$0.5) redshifts using AGN-fuelled radio galaxies. 

Full details of the sample selection, observations and data reduction of the TONS08 survey can be found in \citet{brand}. TONS12, TONS16w and TONS08w were all selected, observed and reduced using the same procedure. The only exception to this is in the selection of the TONS08w sample: sample members were selected by only considering candidates with an angular distance of $\le$10 arcsec between the optical and radio positions. This sample may therefore suffer from some incompletenesses.

\begin{table*}
\begin{center}
\begin{tabular}{rrrrrrl}
\hline\hline
Survey & Type & radio limits & optical limits & colour cut & area (sr.) & ref. \\
\hline
TONS08 & $ROz$ & $s_{1.4}>$3 & $E<$19.83 & $B-R >$1.8 & 0.00688 & {\scriptsize \citet{brand}}\\
TONS12  & $ROz$ & $s_{1.4}>$3 & $E<$19.64 & $B-R >$1.8 & 0.00744 & {\scriptsize this paper}\\
TONS16w & $ROz$ & $s_{1.4}>$30 & $E<$19.5  & $B-R >$1.8 & 0.0151 & {\scriptsize this paper}\\
TONS08w & $ROz$ & $s_{1.4}>$30 & $E<$19.5  & $B-R >$1.8 & 0.02786 &{\scriptsize this paper}\\
Lacy    & $ROz$ & $s_{1.4}>$20 &17$<E<$20.2& APM$^{5}$ &0.0122&{\scriptsize \citet{lac}}\\
TOOT    & $Rz$  & $s_{151}>$100 & NONE               &NONE&& {\scriptsize \citet{hr}}\\
7CII    & $Rz$  & $s_{151}>$500 & NONE               &NONE&& {\scriptsize\citet{wil02}}\\
Sadler  & $ROz$ & $s_{1.4}>$2.8 &14$<B_j<$19.4& AGN$^{1}$ &0.099& {\scriptsize\citet{sad}}\\
2dF     & $Oz$  & NONE          & $B_j<$19.4 & NONE & 0.61 & {\scriptsize\citet{col}} \\
NVSS$^{2}$ & $R$& $s_{1.4}>$2.5  & NONE & NONE &10.36 & {\scriptsize\citet{con}}\\
FIRST$^{3}$& $R$& $s_{1.4}>$2  & NONE & NONE & 2.61 & {\scriptsize\citet{bwh}}\\
PN & $Rz$ & $s_{1.4}>$500 & NONE$^{4}$ & NONE & 9.3 & {\scriptsize\citet{pn}}\\
\hline\hline
\end{tabular}
{\caption[Surveys]{\label{tab:samples} Table summarising the various samples used in and/or referred to in this paper. $R$ denotes radio galaxy surveys, $O$ denotes surveys with an optical magnitude cut and $z$ denotes spectroscopic redshift surveys. The radio limits are in mJy. Notes: 1. AGN were identified using Principal Component Analysis (PCA) applied to the spectra \citep{fol}. 2. NVSS is 50 per cent complete at 2.5 mJy and 99 per cent complete at 3.5 mJy. 3. FIRST is 95 per cent complete at the given limiting flux densities. 4. \citet{pn} impose a redshift limit of $z$=0.1. 5. Galaxies were identified from APM charts.  
}}
 \end{center}
 \end{table*}

The selection criteria for all TONS surveys were based on cross-matching positions of objects in radio and optical surveys. An initial selection was made of NVSS targets \citep{con} in the chosen area of sky with 1.4 GHz flux densities, $s_{1.4}>$3 mJy. The NVSS positions were matched with APM positions \citep{mcm} of objects with $E<$19.5. We selected any objects with APM and NVSS positions with an offset of $\le$ 20 arcsec from each other. We then plotted radio contours from the FIRST survey \citep{bwh} over optical POSS-II images and overplotted the positions of the APM and NVSS objects, identifying real identifications by eye. For full details of this technique we refer the reader to \citet{brand}.

To correct for plate-to-plate variations in the magnitudes obtained from the APM survey, we modified the $O$ and $E$ magnitudes using corrections derived from comparison of the APM magnitudes to GSC-2 magnitudes for each POSS-II plate (R. White priv. com.). We modified the magnitude cut from our original target of 19.5 to 19.83 and 19.64 in the TONS08 and TONS12 sub-sample respectively (the magnitude corresponding to the new limit of the most shallow plate). All following analysis uses these corrected values. 

The final selection criteria was that $O-E\ge$1.8. This cuts out all the bluer objects which tend to be stars (which by coincidence lie close to the line of sight to the radio source), quasars (which can lie at much higher redshifts) and star-burst galaxies (which tend to be at low redshifts due to their low radio luminosity).

Table.~\ref{tab:the_samples} summarises the total number of objects in each of the TONS sub-samples as well as the predicted number from a model redshift distribution described in Sec.~\ref{sec:nz}.

\begin{table}
\begin{center}
\begin{tabular}{rrr}
\hline\hline
sample & total number & predicted number\\
\hline
TONS08                    & 84 & 95.1$\pm4.5$ \\
TONS12                    & 107& 97.6$\pm4.5$ \\
TONS08w                   & 47 & 74.8$\pm3.5$ \\
%TONS08w$\textunderscore$ex& 37 & 56.4$\pm2.6$  \\
TONS16w                   & 40 & 40.6$\pm1.9$  \\
\citet{lac}               & 36 & 47.5$\pm2.7$  \\
\citet{sad}               & 433&428.0$\pm27.9$ \\
TONS0812                 & 191 &192.7 $\pm6.4$\\
TONS0812$\textunderscore$SS              & 92 &46.9 $\pm3.4$\\
TONS0812$\textunderscore$nonSS           & 99 &145.8$\pm5.4$\\
\hline\hline
\end{tabular}
{\caption[]{\label{tab:the_samples} The total number of radio galaxies in each of the TONS sub-samples compared to that predicted by the model redshift distribution. TONS0812 is the combined TONS08 and TONS12 sub-samples, TONS0812$\textunderscore$SS and TONS0812$\textunderscore$nonSS are the combined TONS08 and TONS12 sub-samples within super-structure regions and out of super-structure region respectively (see Sec.~\ref{sec:ssreg}).
}}
 \end{center}
 \end{table}

\section{OBSERVATIONS}
\label{sec:tons_obs}

Optical spectra were obtained during the period October 2000 - May 2003 on the 2.6m Nordic Optical Telescope (NOT) using the Andalucia faint object spectrograph, the 4.2m William Herschel telescope (WHT) using ISIS, the 2.7m Smith reflector at McDonald with the Imaging grism instrument (IGI) \citep{hill02}, and the Hobby-Eberly telescope (HET) using the Marcario low resolution spectrograph (LRS) \citep{hill98}. 
 
%Exposures of approximately 10-30 minutes were taken (depending on the magnitude of the source, the weather conditions and the telescope used). The position angle of the slit on the sky was either oriented to include more than one possible identification, or along the radio axis if the object was extended or at parallax.

%\begin{table*}
%\begin{center}
%\begin{tabular}{rrrrr}
%\hline\hline
%Telescope & Spectrograph & Grating & Dispersion & Wavelength coverage \\
%\hline
%WHT & ISIS (red) & R158R& 2.9 & 5500-9000\\
%WHT & ISIS (blue)& R158B& 1.62&3000-5500\\
%NOT & ALFOSC& No.4  & 8.1 & 3200-9100\\
%2.7 & IGI   &  & 6.4 &3900-9500\\
%HET & LRS   & G2 & 2.0 & 4280-7278\\
%\hline\hline
%\end{tabular}
%{\caption[Surveys]{\label{tab:instr} Table summarising the spectrographs used on the different telescopes and their instrumental dispersion ($\rm \AA$/pixel) and wavelength coverage ($\rm \AA$).
%}}
% \end{center}
% \end{table*}

\begin{figure}
\begin{center}
\setlength{\unitlength}{1mm}
\begin{picture}(50,35)
\put(70,-5){\includegraphics{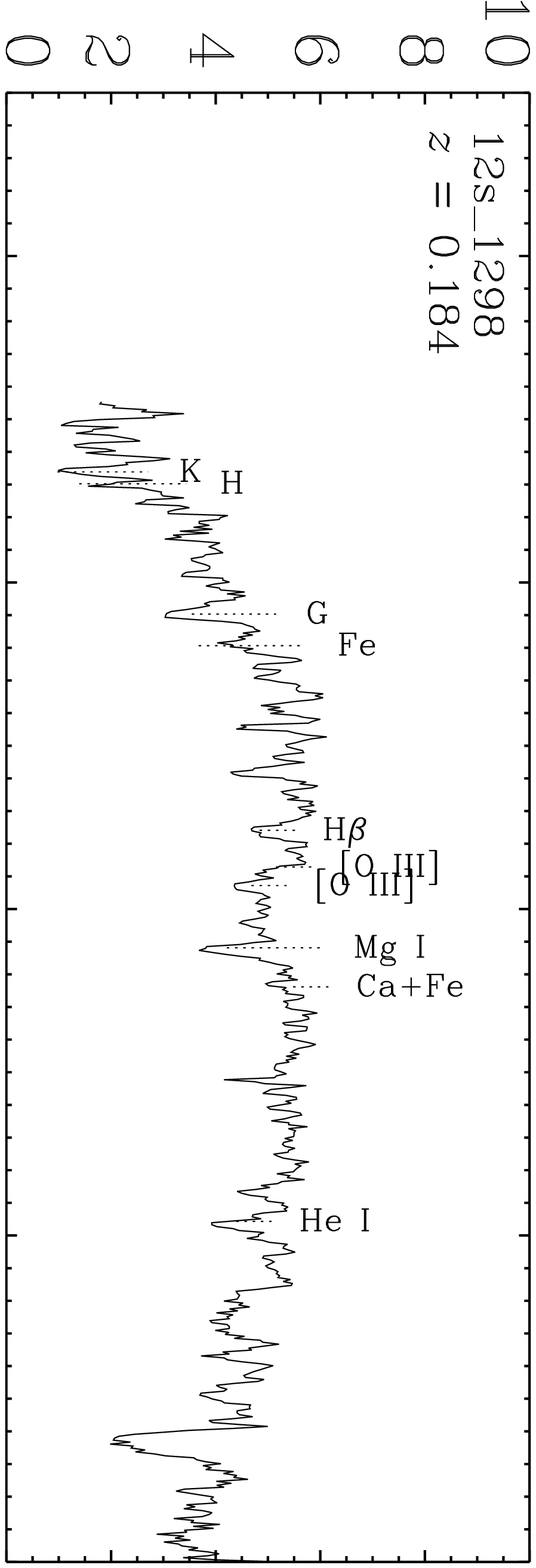}}
\end{picture}
\end{center}
{\caption[junk]{\label{fig:12spectra} Spectra of all the radio galaxies in TONS12 that are not in TOOT, reported by Benn (priv. comm.) and/or do not have redshifts from the literature. See Table.~\ref{tab:12summary} for details. The Flux density is in units of 10$^{-20}\ \rm{W}\rm{m}^{-2}\rm{\AA}^{-1}$. Shown here is an example spectrum. Plots of all the spectra can be found in the version to be published in MNRAS in early 2005.
}}
\end{figure}

\begin{figure}
\begin{center}
\setlength{\unitlength}{1mm}
\begin{picture}(50,35)
\put(70,-5){\includegraphics{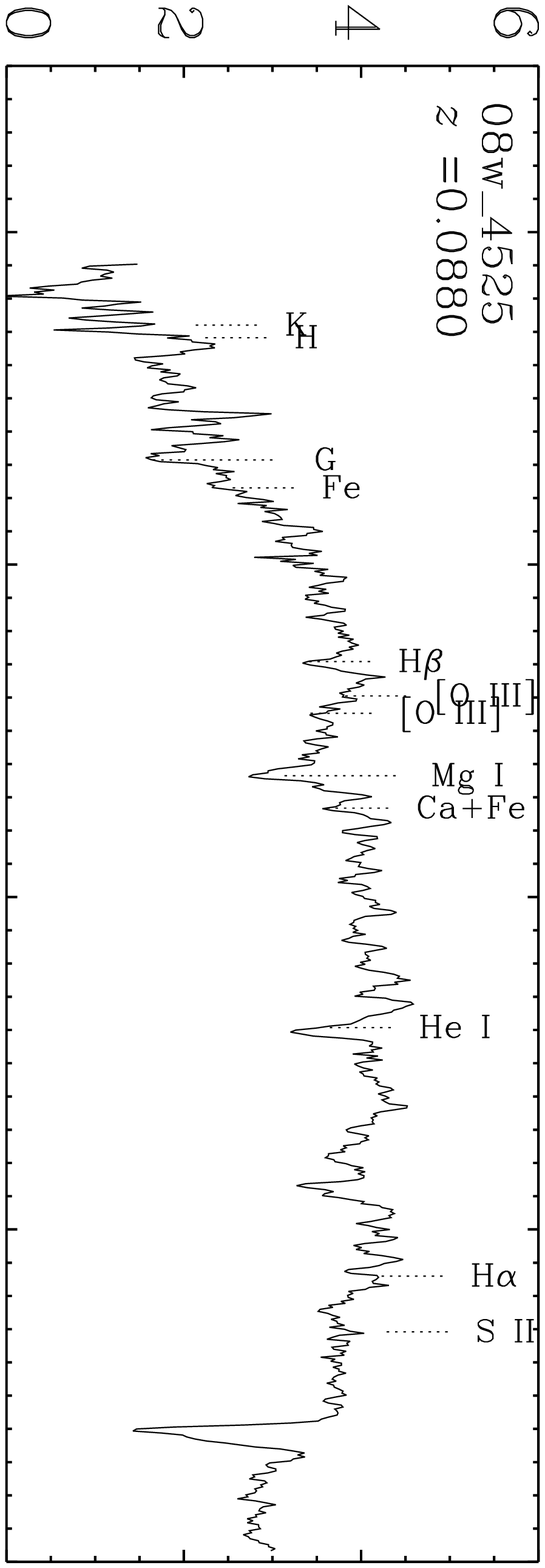}}
\end{picture}
\end{center}
{\caption[junk]{\label{fig:08wspectra} Spectra of all the radio galaxies in TONS08w that are not in TONS08, TOOT08 and/or not not have redshifts from the literature. See Table.~\ref{tab:08wsummary} for details. The flux density is in units of 10$^{-20}\ \rm{W}\rm{m}^{-2}\rm{\AA}^{-1}$. Shown here is an example spectrum. Plots of all the spectra can be found in the version to be published in MNRAS in early 2005.
}}
\end{figure}

\begin{figure}
\begin{center}
\setlength{\unitlength}{1mm}
\begin{picture}(50,35)
\put(70,-5){\includegraphics{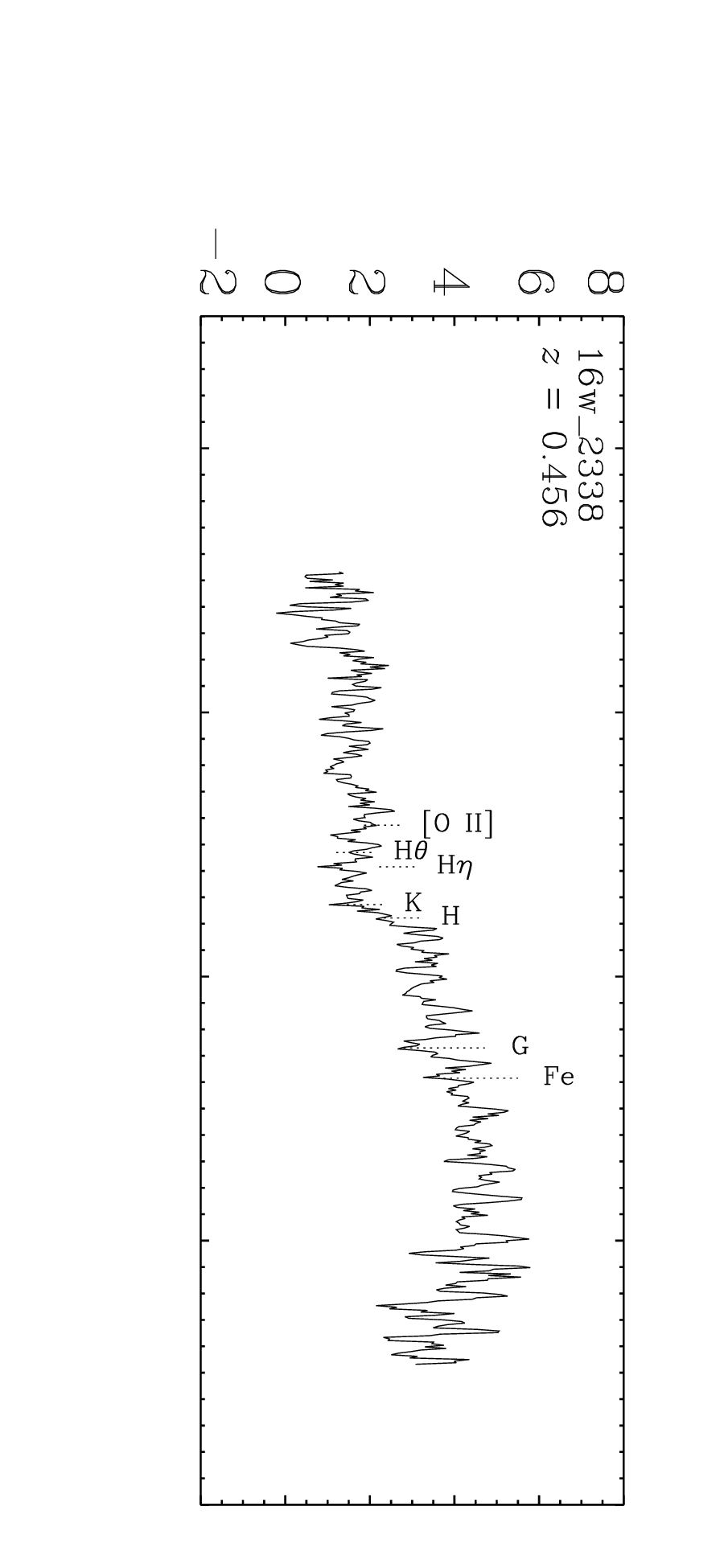}}
\end{picture}
\end{center}
{\caption[junk]{\label{fig:16spectra} Spectra of all the radio galaxies in TONS16 that are not in TOOT16 and/or do not have redshifts from the literature. See Table.~\ref{tab:16summary} for details. The flux density is in units of 10 $^{-20}\ \rm{W}\rm{m}^{-2}\rm{\AA}^{-1}$. Shown here is an example spectrum. Plots of all the spectra can be found in the version to be published in MNRAS in early 2005.
}}
\end{figure}

The spectra were reduced using standard {\small IRAF} procedures. Spectra and redshifts for the TONS08 sub-sample have been presented in \citet{brand}. Spectra for the observed radio galaxies (with no previously known redshift) for the TONS12, TONS08w and TONS16w sub-samples are shown in Fig.~\ref{fig:12spectra}, Fig.~\ref{fig:08wspectra} and Fig.~\ref{fig:16spectra} respectively. As found for the TONS08 sub-sample, the vast majority of objects in the entire TONS survey were identified as moderate redshift radio galaxies.
All spectra are smoothed for presentation purposes and sorted by redshift for clarity. In most cases, redshifts were determined from absorption lines (very few objects exhibited emission lines). The estimated redshifts are presented in Table.~\ref{tab:12summary}, Table.~\ref{tab:08wsummary} and Table.~\ref{tab:16summary}. These were checked by performing a cross-correlation between the spectra and both a rest frame composite spectrum obtained from the TONS08 data \citep{brand} and an evolved population GISSEL model \citep{bc}.  

Of all the sub-samples, only TONS16w overlaps with the second data release of the Sloan Digital Sky Survey (SDSS; \citealt{aba04}). 11/40 of the TONS16w sample have spectroscopic redshifts in SDSS. In all cases, the redshifts agree to within 0.003. 

\begin{table*}
%\begin{figure*}
\vbox to400mm{\vfil 
\begin{picture}(100,100)
\put(460,-65){\includegraphics{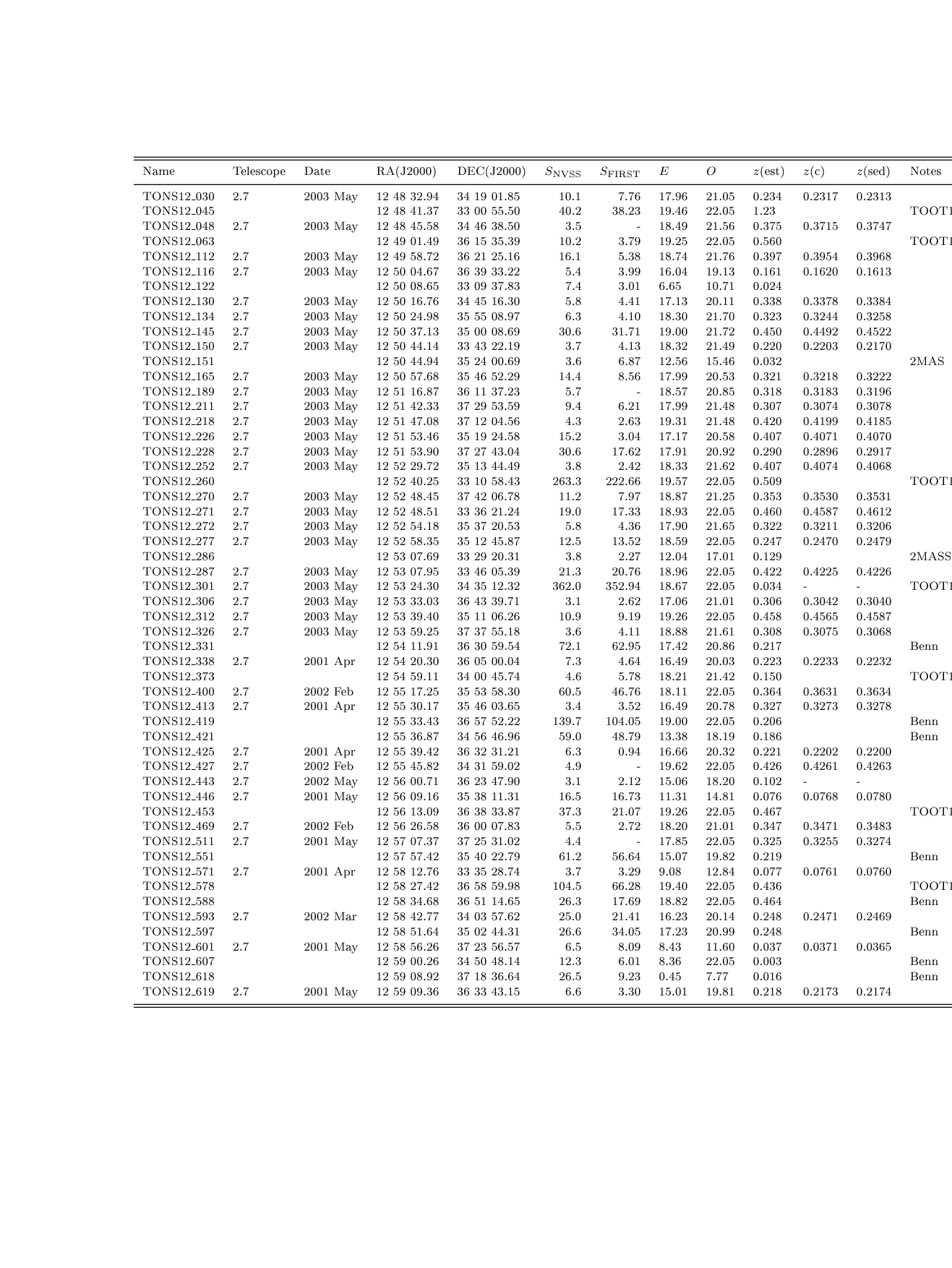}}
\end{picture}
{\caption[Table 1]{\label{tab:12summary} A summary of key information 
on the TONS12 sample. Note that the RA and DEC are that of the radio positions except in the case of multiple sources where the positions are from APM.  $S_{\rm{NVSS}}$ and $S_{\rm{FIRST}}$ are the integrated radio flux densities (mJy) in the NVSS and FIRST catalogues respectively. In cases where FIRST resolves out multiple components, $S_{\rm{FIRST}}$ is the total integrated flux density. The $O$ and $E$ magnitudes are the APM corrected magnitudes. $z$(est) is the redshift estimated by identifying lines by eye and calculating the mean redshift. $z$(c) and $z$(sed) are the redshifts obtained by cross-correlating the spectra with the combined de-redshifted spectra and the GISSEL model respectively. The notes denote the name of the object in other samples. A - denotes that the cross-correlation was unsuccessful. Spectra of the TOOT objects will be presented in Rawlings et al. (in prep). Spectra denoted by Benn are from Benn (Priv. Comm.). Other redshifts were obtained from the literature.

}}
\vfil}
%\end{figure*}

\end{table*}

%\clearpage

\addtocounter{table}{-1}

\begin{table*}
%\begin{figure*}
\vbox to400mm{\vfil 
\begin{picture}(100,100)
\put(460,-65){\includegraphics{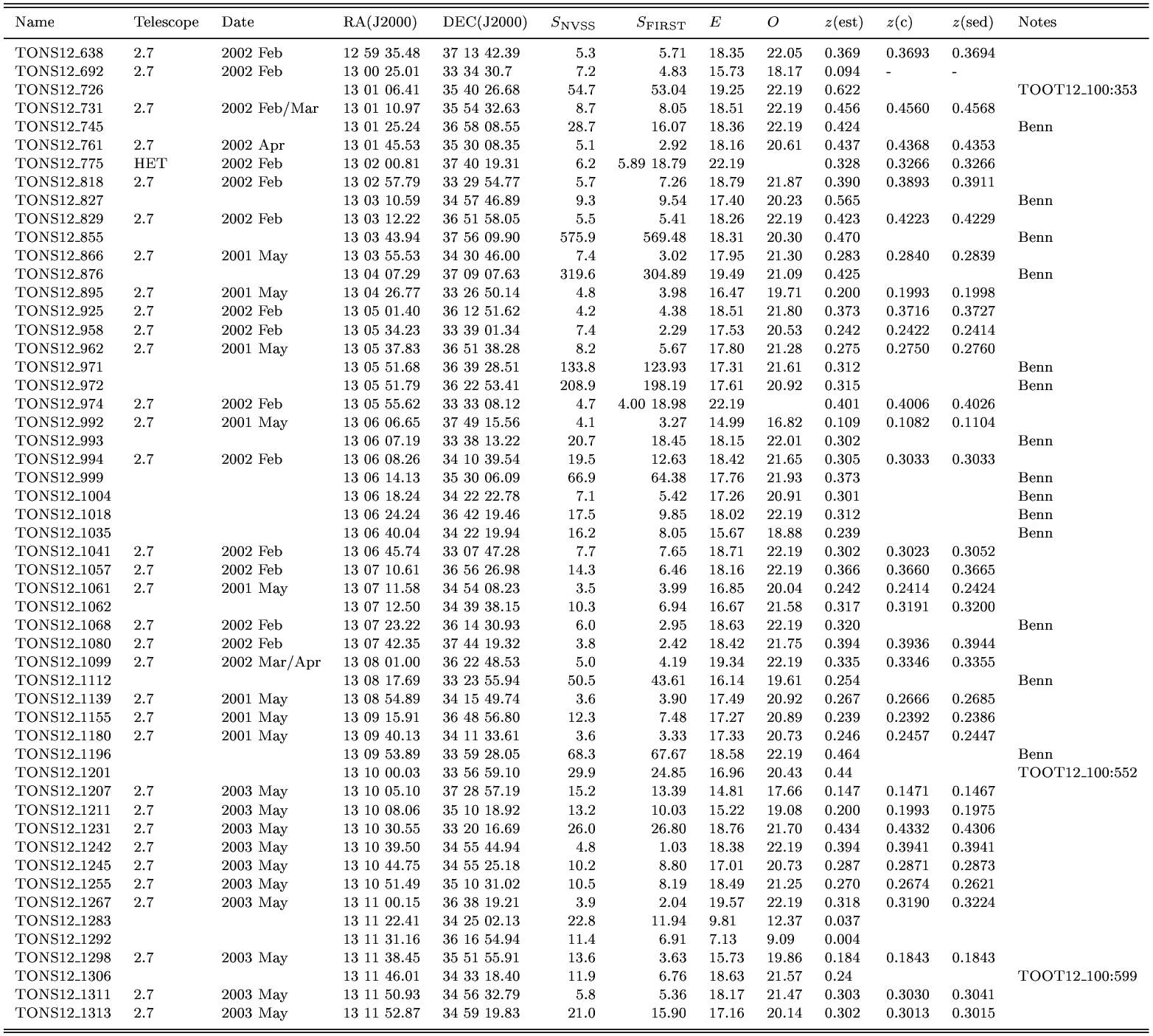}}
\end{picture}
{\caption{\bf (cont).}}
\vfil}
%\label{landfig}
%\end{figure*}
\end{table*}

\begin{table*}
%\begin{figure*}
\vbox to400mm{\vfil 
\begin{picture}(100,100)
\put(460,-65){\includegraphics{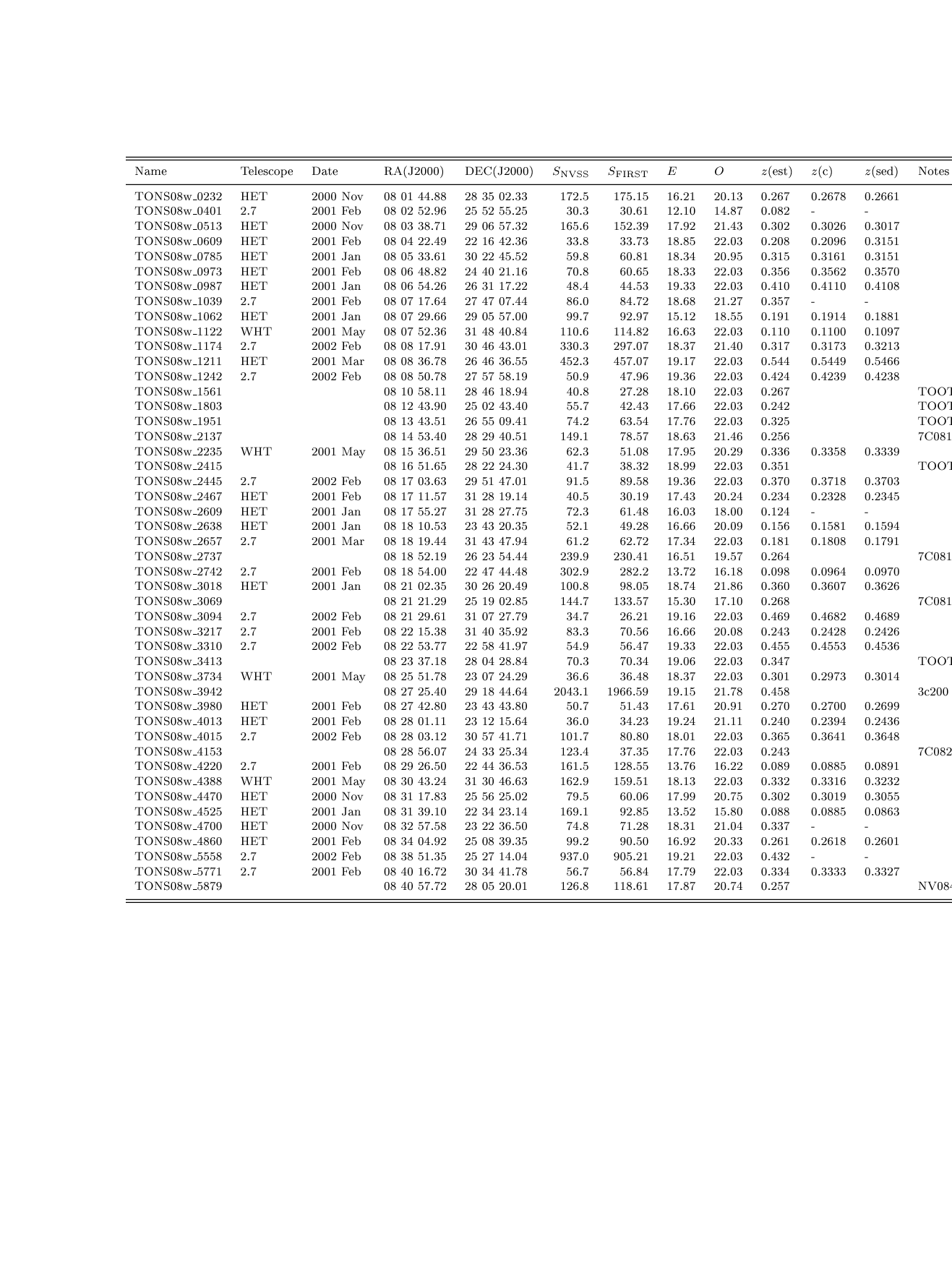}}
\end{picture}
{\caption[Table 1]{\label{tab:08wsummary} A summary of key information 
on the TONS08w sample. Note that the RA and DEC are that of the radio positions except in the case of multiple sources where the positions are from APM.  $S_{\rm{NVSS}}$ and $S_{\rm{FIRST}}$ are the integrated radio flux densities (mJy) in the NVSS and FIRST catalogues respectively. In cases where FIRST resolves out multiple components, $S_{\rm{FIRST}}$ is the total integrated flux density. The $O$ and $E$ magnitudes are the APM magnitudes. $z$(est) is the redshift estimated by identifying lines by eye and calculating the mean redshift. $z$(c) and $z$(sed) are the redshifts obtained by cross-correlating the spectra with the combined de-redshifted spectra and the GISSEL model respectively. The notes denote the name of the object in other samples. - denotes that the cross-correlation was unsuccessful. Redshifts with no observations were obtained from the literature.

}}
\vfil}
%\end{figure*}

\end{table*}

\begin{table*}
%\begin{figure*}
\vbox to400mm{\vfil 
\begin{picture}(100,100)
\put(460,-65){\includegraphics{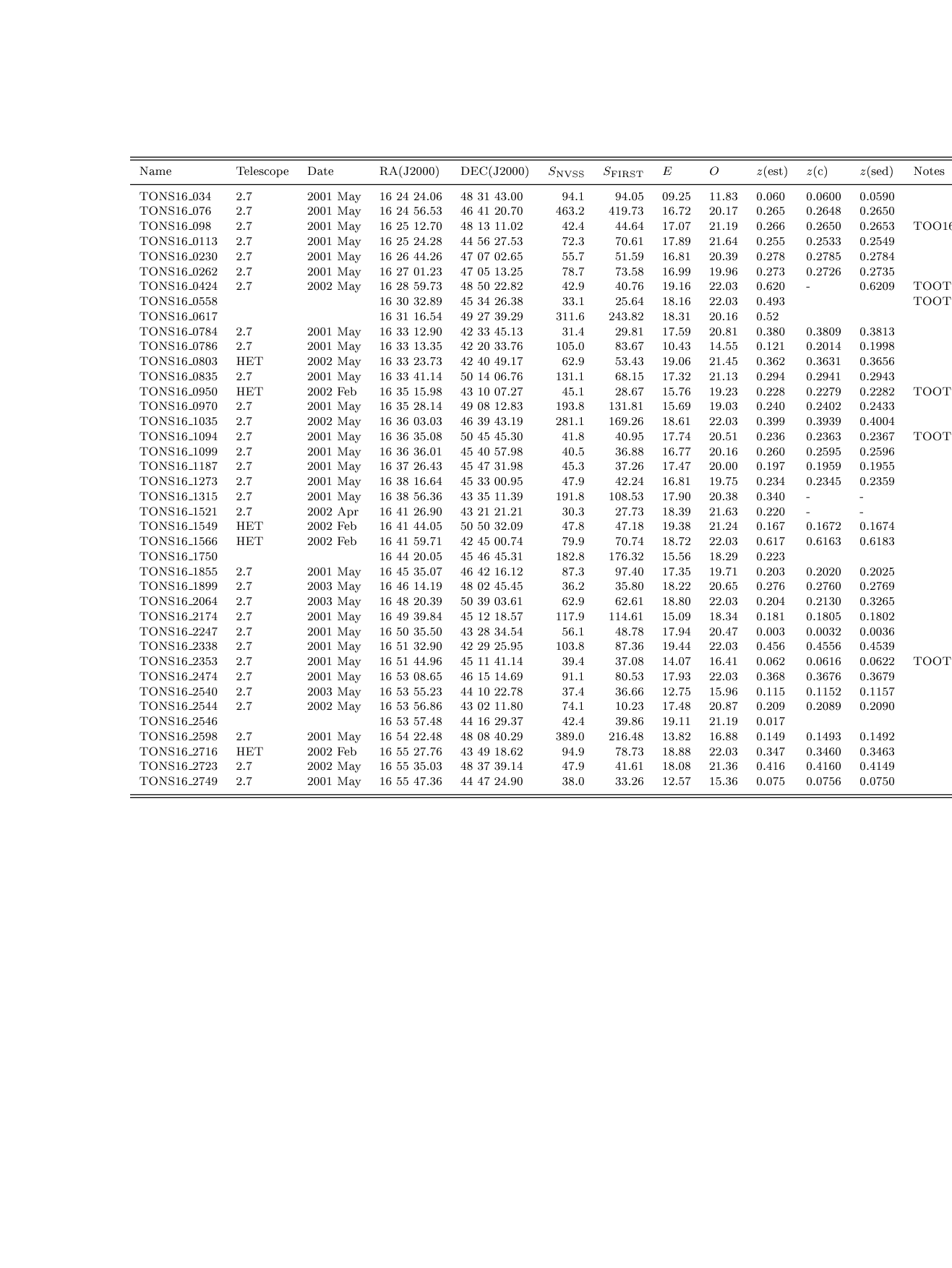}}
\end{picture}
{\caption[Table 1]{\label{tab:16summary} A summary of key information 
on the TONS16 sample. Note that the RA and DEC are that of the radio positions except in the case of multiple sources where the positions are from APM.  $S_{\rm{NVSS}}$ and $S_{\rm{FIRST}}$ are the integrated radio flux densities (mJy) in the NVSS and FIRST catalogues respectively. In cases where FIRST resolves out multiple components, $S_{\rm{FIRST}}$ is the total integrated flux density. The $O$ and $E$ magnitudes are the APM magnitudes. $z$(est) is the redshift estimated by identifying lines by eye and calculating the mean redshift. $z$(c) and $z$(sed) are the redshifts obtained by cross-correlating the spectra with the combined de-redshifted spectra and the GISSEL model respectively. The notes denote the name of the object in other samples. - denotes that the cross-correlation was unsuccessful. Spectra of the TOOT objects will be presented in Rawlings et al. (in prep). Other redshifts were obtained from the literature.

}}
\vfil}
%\end{figure*}

\end{table*}

\section{THE TONS REDSHIFT DISTRIBUTIONS}

\subsection{Obtaining a universal N(z)}
\label{sec:nz}

To perform clustering analysis on a survey, it is crucial to know the underlying redshift distribution of the population under study. This is often obtained by simply using a redshift distribution that is either smoothed or fitted to the data itself (e.g. \citealt{ste}). However, the obvious presence of large-scale structure in our sub-samples means that this will not be representative of the true underlying redshift distribution. 

We required a model redshift distribution derived from a data set over a sufficiently large area to reduce the effects of large-scale structure and which we could modify for different optical magnitude and radio flux density limits of each of our sub-samples. There are currently no model redshift distributions derived from surveys going down to NVSS / FIRST flux densities in the literature. \citet{dp} and \citet{wil01} model the radio luminosity function of relatively bright ($s_{\rm{1.4}}>$100 mJy) radio galaxies from which a redshift distribution can be extrapolated to lower flux densities and the correct radio frequency. However these models can display unphysical spikes at low redshifts and the extrapolation results in large uncertainties. These models were not accurate enough for our purposes.

Instead, we used a maximum likelihood technique \citep{mar} to fit our own bivariate (radio and optical) luminosity function (BLF) to radio galaxies in the 2dF galaxy redshift survey \citep{sad}. These data are over a sufficiently large area to smooth over any large-scale structure. The BLF can be integrated over any optical apparent magnitude and radio flux density limits to obtain a model redshift distribution for each of the TONS sub-samples. \citet{brand} described the form of the bivariate luminosity function that was modelled and the best fit parameters of this model. Note that because to TONS survey extends out only to moderate redshift, the redshift evolution is not well constrained. \citet{cj} found that the lower luminosity (FR I dominated) AGN population is consistent with a constant co-moving space density with redshift. Fixing the BLF to a model with no evolution with redshift, resulted in no significant difference in the model redshift distribution. 

\noindent Because the TONS survey optical apparent magnitude limits were defined in the $E(\approx R)$ band but the BLF is defined for the $b_j$ band, we converted the apparent $R$-band magnitude limits to absolute $R$ magnitude for each redshift and applied a colour correction of $B-R$=1.52 (calculated from the difference in flux in the rest-frame of a template SED of an evolved stellar population \citep{bc} measured in the 2 different bands). We assumed a colour correction of $B_j-B=-$0.11 \citep{fre}.

\begin{figure}
\begin{center}
\setlength{\unitlength}{1mm}
\begin{picture}(150,135)
\put(-2,60){\includegraphics{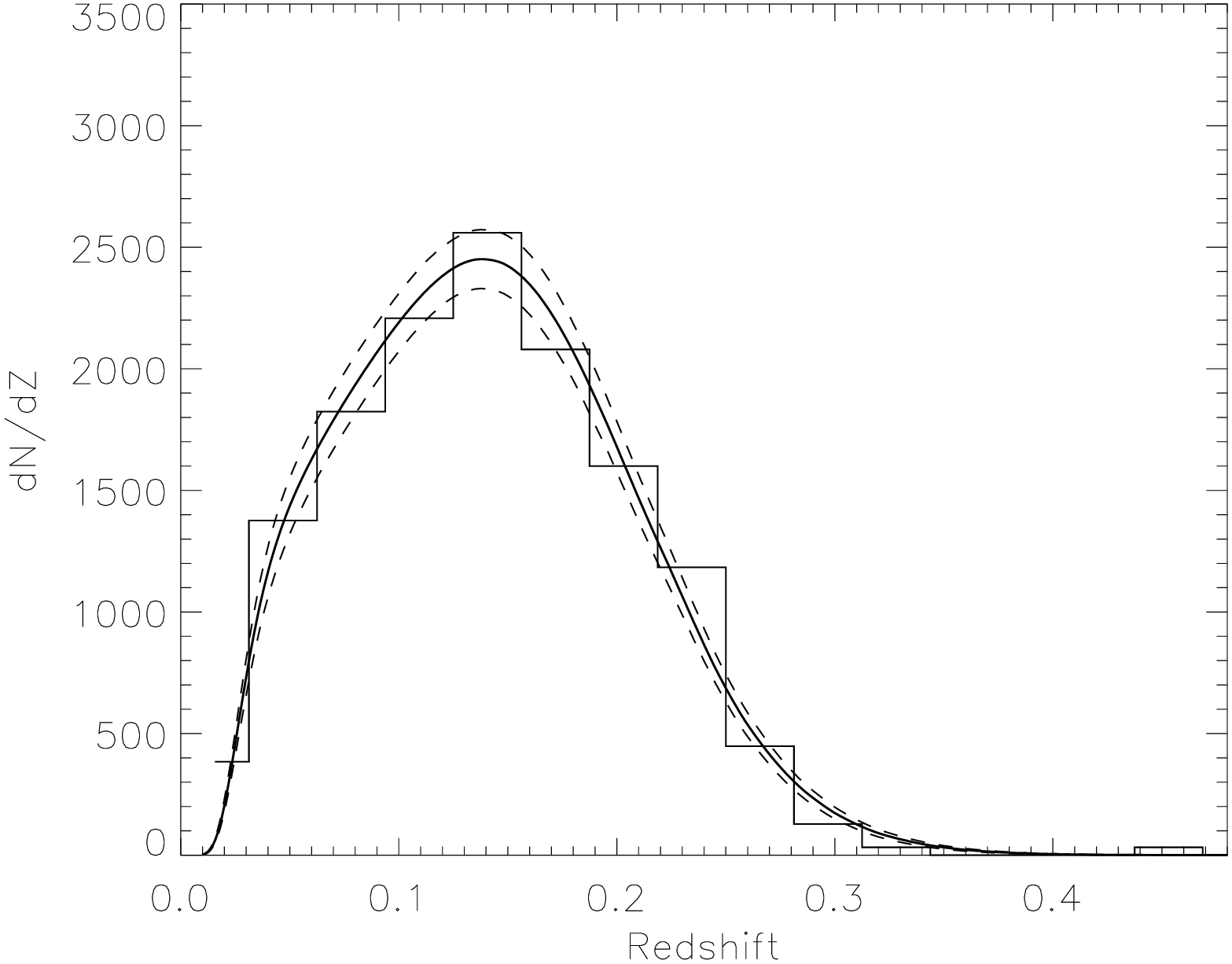}}
\put(-2,-10){\includegraphics{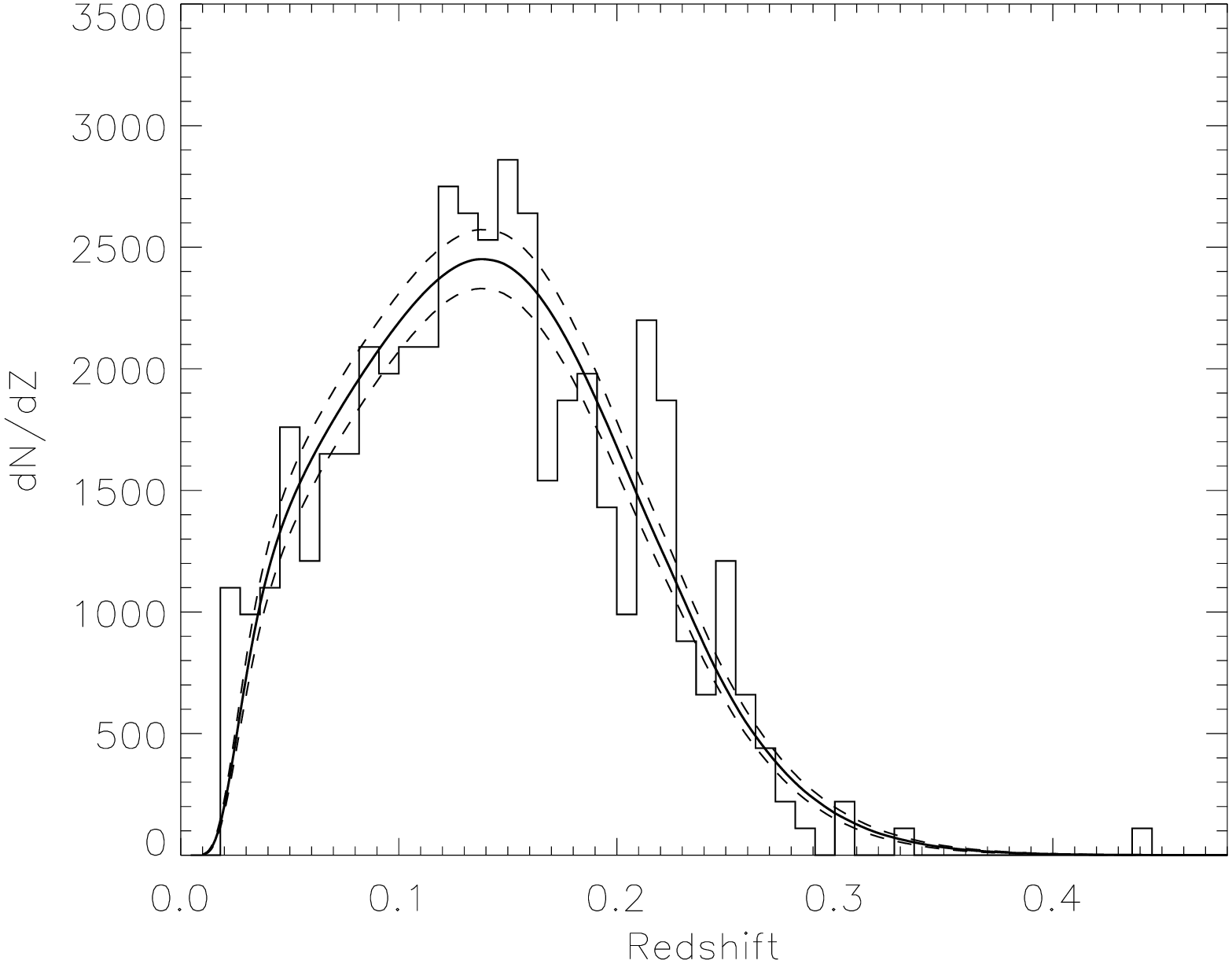}}
\end{picture}
\end{center}
{\caption[junk]{\label{fig:sad} The redshift distribution of the \citet{sad} sample with the model redshift distribution overplotted (solid line). The $\pm 1\sigma$ errors on the model are overplotted (dashed lines). We plot the distribution using 16 and 55 redshift bins between $z$=0 and $z$=0.5 for the top and bottom figures respectively. Even with 433 radio galaxies over a larger area, the redshift distribution is not smooth.
}}
\end{figure}

The redshift distribution of the \citet{sad} radio galaxies is shown is Fig.~\ref{fig:sad}. Overplotted is the fitted model redshift distribution. We note that although this survey is is over 13 times the area of TONS08, it still exhibits features due to large-scale structure and this may bias the resulting fit of the BLF. We minimised this effect by effectively smoothing the redshift distribution (by modifying the redshifts of each radio galaxy by an amount drawn from a Gaussian distribution of FWHM 0.05).

\subsection{Errors}

We found the errors on the redshift distribution using the following method to allow for the deviations in the derived model parameters. We split the parameter space into discrete grid spaces. For each grid space, we calculated the maximum likelihood, $S$ of this parameter combination and converted it to a probability ($P\propto \exp(-S/2)$). We then normalised the probabilities so that the total probability in all grid spaces added up to one and converted to a cumulative probability. We calculated the number of radio galaxies expected for each redshift interval given the survey limits and the parameter values in that box. We then weighted this by multiplying by the probability of the parameter combination. By adding up the weighted number of galaxies for each redshift interval, we should have obtained a very similar redshift distribution to that obtained by the best fit parameter values. The advantage of this procedure is that if there is more than one parameter combination which fits the BLF well, these will also have an influence on the final redshift distribution. 

To calculate the errors, we performed a Monte-Carlo simulation in which we chose 5000 random numbers between 0 and 1. Because the cumulative probabilities for all parameter space grid spaces were normalised to run from 0 to 1, each random number could be associated with a unique parameter combination. Those parameter combinations which have higher relative probabilities occupy a larger proportion of this probability space and will therefore have had a greater chance of being selected. The redshift distribution was then calculated for all parameter combinations that were selected. The minimum and maximum $\pm$1$\sigma$ errors at each redshift interval were calculated by determining where 67 per cent of the measurements fell. 

It is likely that other quantities will also have contributed to the error in the measurement of the redshift distribution. Errors may have been introduced by incompletenesses in both the radio and optical surveys at faint flux densities: The NVSS catalogue is $\approx$ 90 per cent complete at $s_{1.4}$=3 mJy \citep{con}. Before correcting for plate-to-plate variations, the APM magnitudes have a global rms uncertainty of 0.5 mag \citep{mcm}. We assumed an error of 0.3 mag, which is the difference between the magnitude corrections of the two POSS-II plates in the TONS08 region. We incorporated this into the total error on each redshift bin by determining the $R$ magnitude limit in the Monte Carlo simulations as a Gaussian distributed value with a standard deviation of 0.3 about the magnitude limit. 

\subsection{A comparison of the TONS sub-samples with their model redshift distributions}

Table.~\ref{tab:the_samples} shows the number of radio galaxies observed and predicted by our model redshift distribution in all the TONS sub-samples. These numbers are significantly different in most cases. However, it is not surprising that there are large variations in these numbers due to the presence of super-structures and voids which occur on similar scales to that of the survey size ($\sim$100 Mpc diameter at $z\sim$0.3).

\begin{figure*}
\begin{center}
\setlength{\unitlength}{1mm}
\begin{picture}(150,130)
\put(-2,60){\includegraphics{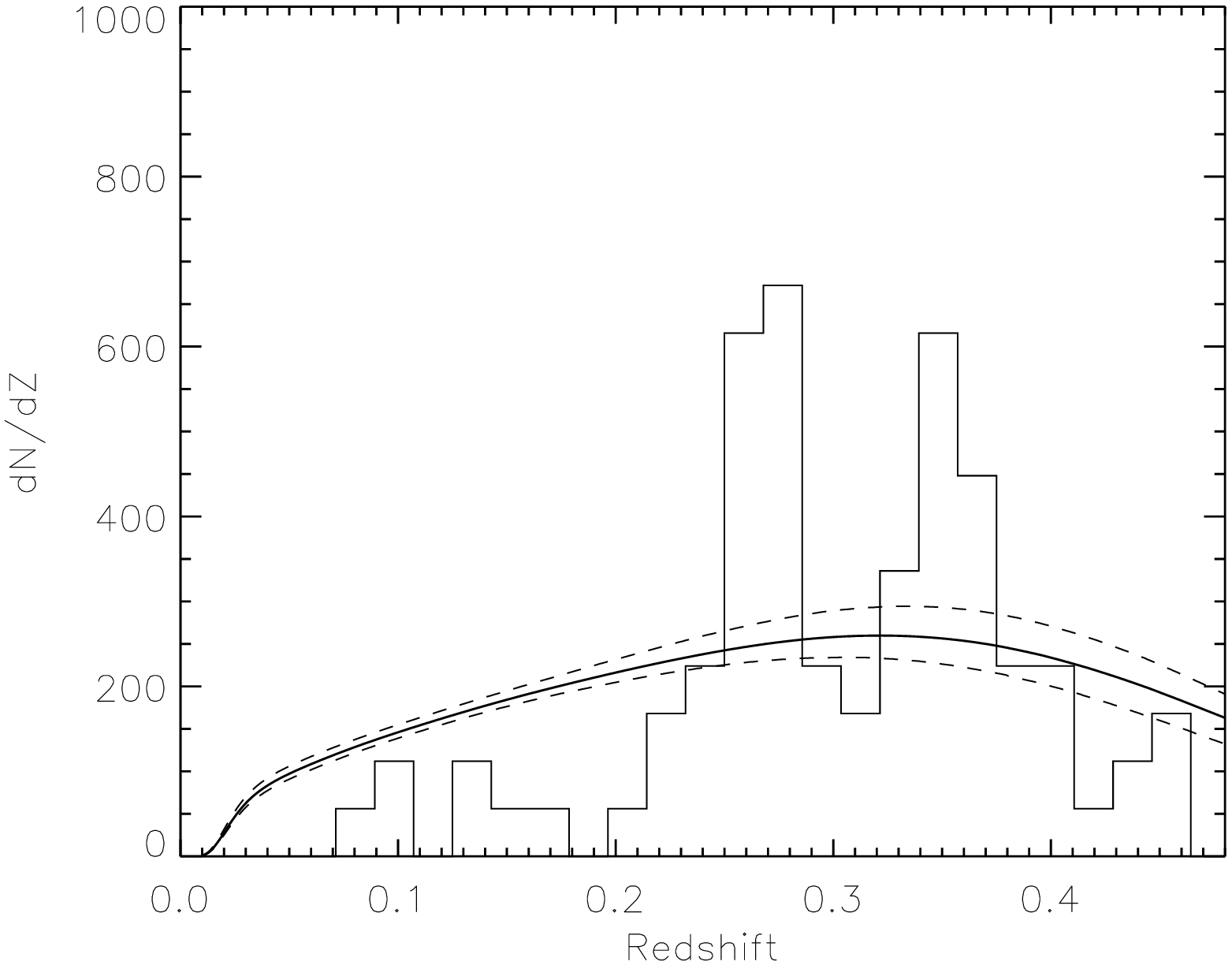}}
\put(70,60){\includegraphics{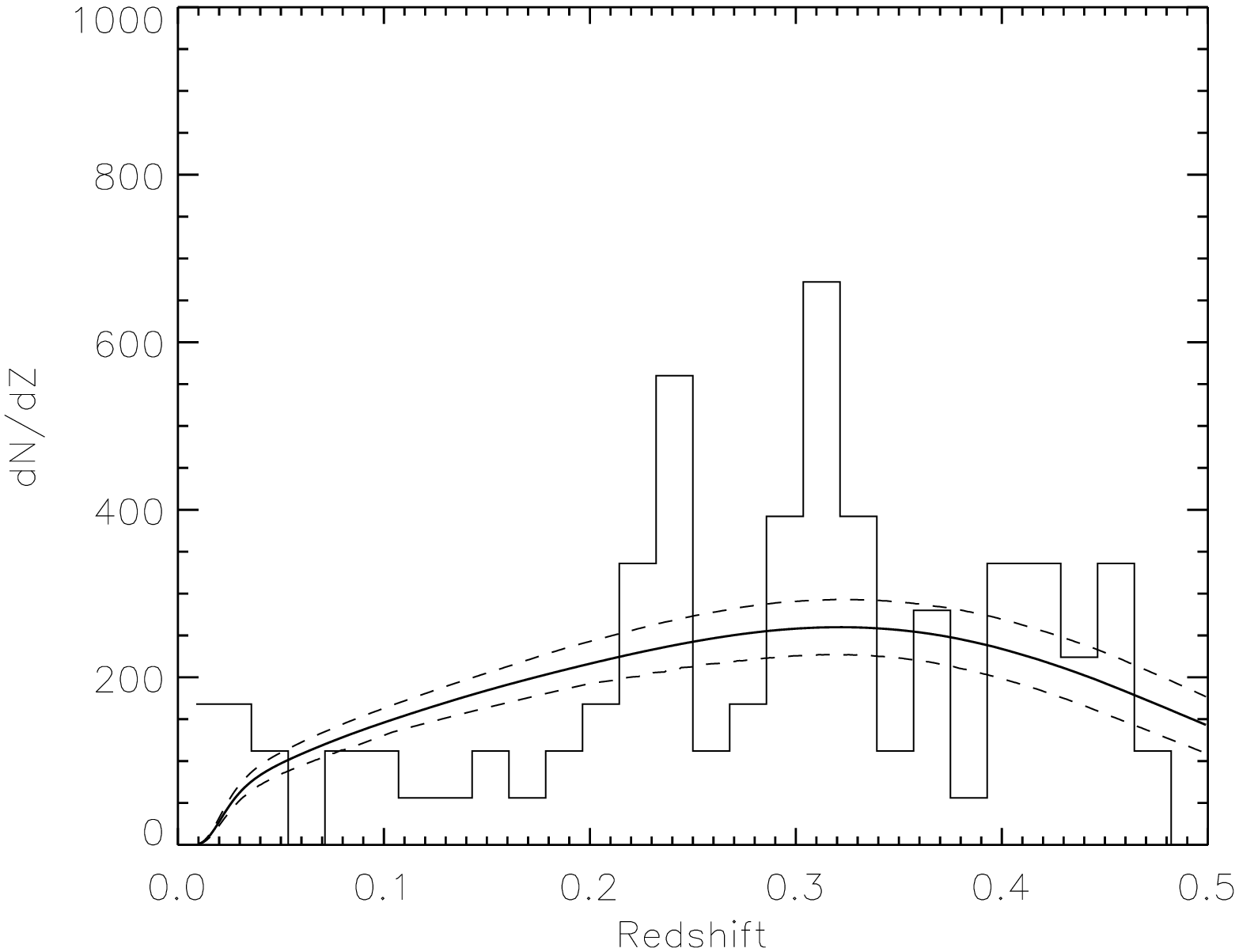}}
\put(-2,-10){\includegraphics{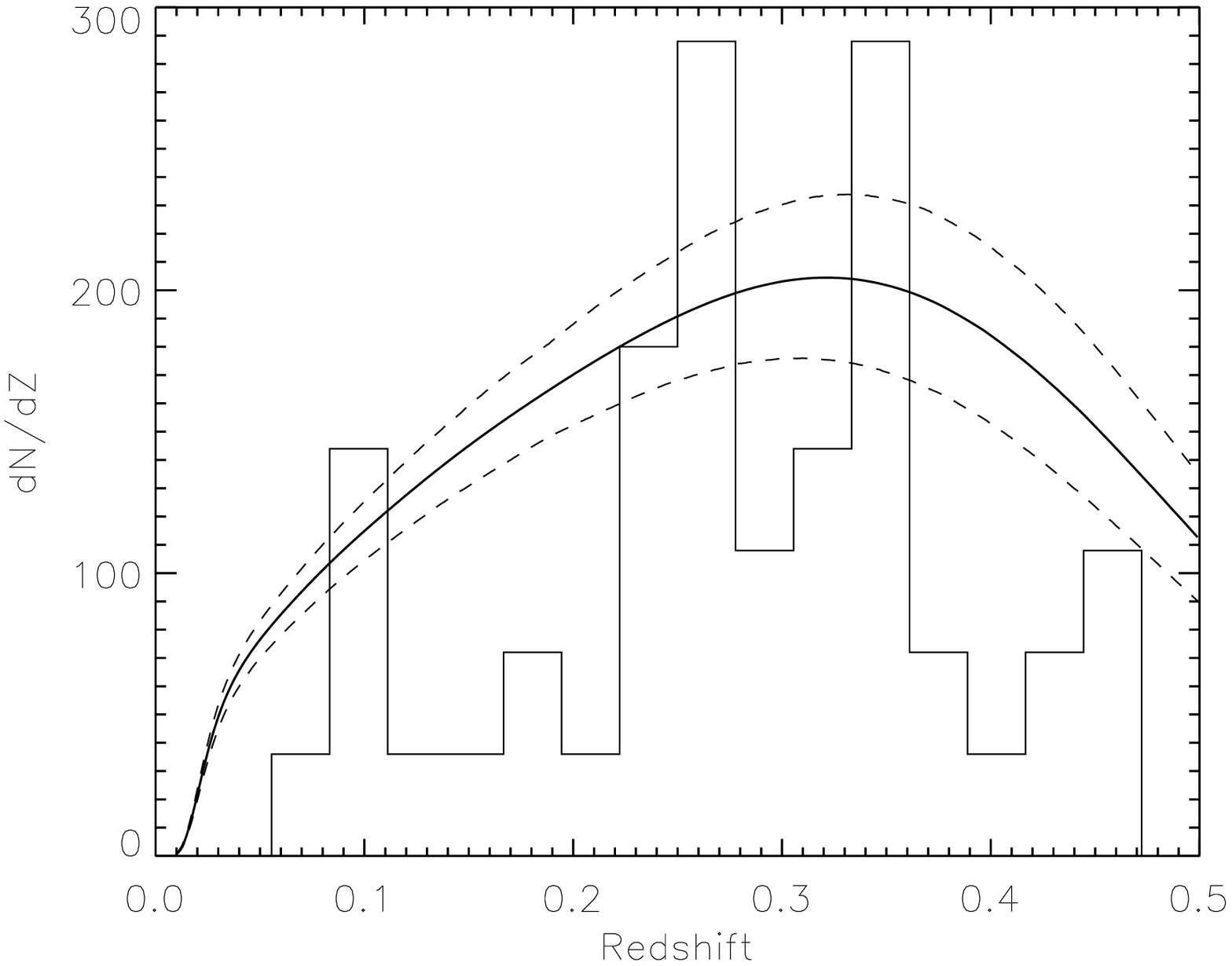}}
\put(70,-10){\includegraphics{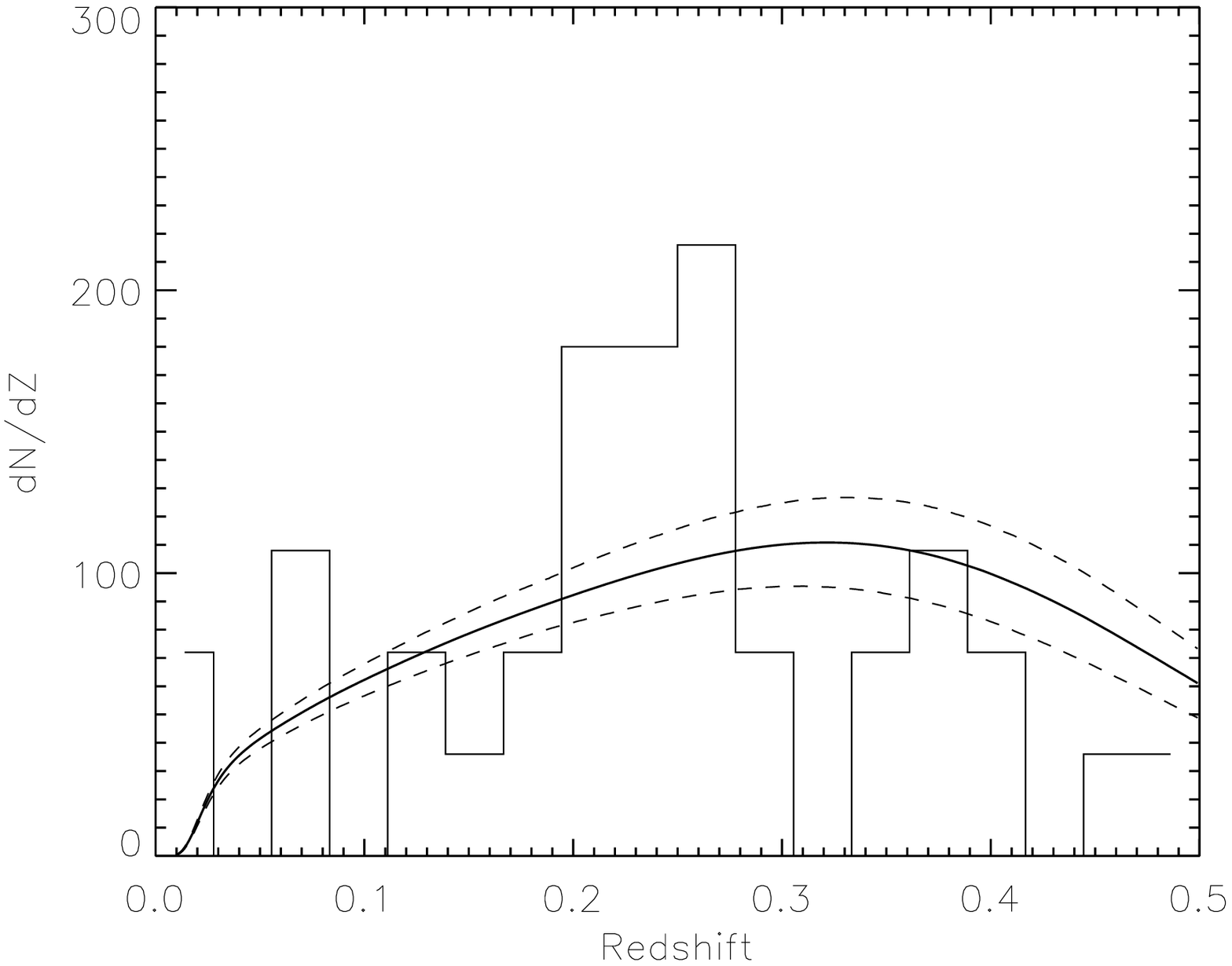}}
\put(28,127){\large\bf TONS08}
\put(100,127){\large\bf TONS12}
\put(28,57){\large\bf TONS08w}
\put(100,57){\large\bf TONS16w}
\end{picture}
\end{center}
{\caption[]{\label{fig:zdist_all} The redshift distribution of the TONS08, TONS12, TONS08w and TONS16w sub-samples (top left to bottom right) with the model redshift distribution overplotted (solid line). The $\pm 1\sigma$ errors on the model are also shown (dashed lines). We plot the distribution using 28 redshift bins between $z$=0 and $z$=0.5.
}}
\end{figure*}

Fig.~\ref{fig:zdist_all} shows the redshift distributions overplotted with the model redshift distributions for the TONS08, TONS12, TONS08w and TONS16w sub-samples. In all cases, large-scale structure has changed the redshift distribution of our sub-sample significantly from that expected for a typical area of sky.

Within the TONS08 sub-sample, there are two prominent spikes at $z\approx$0.27 and $z\approx$0.35 with, perhaps, void-like regions at lower redshifts, accounting for the low overall number of radio galaxies. Analysis of these redshift spikes; their significance and their implications for structure formation theories and/or radio galaxy bias is performed in \citet{brand}. The TONS12 survey covers a similar area to that of TONS08 but in a different region of sky. It too shows two redshift peaks above that expected from the model redshift distribution (but at redshifts $z\approx$0.23 and $z\approx$0.31). That redshift spikes are found in both sub-samples is important as it demonstrates that redshift spikes are probably a universal feature of radio galaxy distributions.

The TONS08w and 16w surveys are wider area samples with higher flux density limits. Within the TONS08w sub-sample, there are fewer radio galaxies than we would expect. However, this may simply be due to incompletenesses in this survey. We may have missed sources with a large offset between the radio and optical positions - something more likely for higher power radio galaxies. Within the TONS16w sub-sample, we observed a similar number of radio galaxies to that predicted from the model redshift distribution. Although there is an excess of radio galaxies at a redshift $z$=0.2-0.3, the number density of radio galaxies is insufficient to determine if there is a significant overdensity. This is perhaps not surprising given the lack of evidence for significant redshift spikes in previous surveys with a similar flux density limit (see \citealt{brand} sec.~5.3).

\section{RADIO GALAXY CLUSTERING}
\label{sec:corr_fn}

\subsection{Calculating the two point correlation function}

The spatial two-point correlation function is the traditional tool for measuring the strength of clustering and hence provides an easy comparison with other surveys. It is also perhaps the most intuitive and simple measure of clustering. It provides fundamental information about the galaxy distribution in the sense that it is the Fourier transform of the power spectrum of the density fluctuations and (when the clustering is still in the linear regime and the bias is known) can be directly related back to fluctuations in the primordial density field. The two-point correlation function measures the excess probability over random d$P$ of finding a pair of objects in two volumes d$V_1$ and d$V_2$ with a separation, $r$ (e.g. \citealt{pn}):

\begin{equation}
\rm{d}P=\rho_{0}^2 [1+\xi(r)] \rm{d}V_{1}\rm{d}V_{2},
\end{equation}

\noindent where $\rho_{0}$ is the average number density. The form of the correlation function is generally well approximated by a power-law:

\begin{equation} \label{eqn:r0}
\xi(r)=\left(\frac{r}{r_0}\right)^{-\gamma},
\end{equation}

\noindent where $r_0$ is the correlation scale length (the separation at which $\xi$=1) and $\gamma$ is the slope of the correlation function and is usually assumed to be equal to 1.8 (e.g. \citealt{gp}). Thus, the strength of any clustering can be estimated by obtaining a value for $r_0$. The larger the value of $r_0$, the greater the clustering signal of the sample. 

In practise, the correlation function is calculated using an estimator. There are several different forms of this estimator, but they all involve counting pairs of objects within the real catalogue, a random catalogue and between the two at different separations, $r$. \citet{ker} find that on small (r$\approx$4.4$h^{-1}$Mpc) scales, the estimators are comparable. However, for larger scales (r$\approx$115$h^{-1}$Mpc), the estimator of \citet{lasz} significantly outperforms the rest:

\begin{equation} \label{eq:xi}
\xi(r)=\frac{DD(r)-2DR(r)+RR(r)}{RR(r)},
\end{equation}
 
\noindent where DD(r) is the number of data-data pairs, RR is the number of random-random pairs and DR is the number of data-random pairs all calculated for separation $r$. This is the estimator that we have used for this analysis.

In order to calculate the two-point correlation function, a set of random catalogues was constructed. These catalogues allow a comparison of the data to randomly distributed objects on the sky. The random catalogues were all constructed with the same number of objects as the real catalogue and the same selection function. They were all assigned a random RA and DEC within the sample boundaries and a random redshift selected from the redshift distribution expected for objects in the survey. We used the redshift distribution calculated in Sec.~\ref{sec:nz}.  The RR and DR pairs used in the calculation of the correlation function were obtained by averaging 1000 simulations of the random catalogue to reduce to negligible any errors due to statistical variations within the random samples.

\subsection{Corrections to the two-point correlation function}

Small survey volumes that are unrepresentative of the overall field must be corrected for the ``integral constraint''. Because the TONS survey sizes are comparable to scales at which the number density of galaxies in the volume can significantly differ from field to field, we need to account for the fact that they are not representative of the general field. Because the correlation function is only calculated for a finite survey volume, in practise the observed number density $n$ is used instead of the global number density $<n>$.
%To understand this, consider a huge volume that has a clustering signal in which the correlation function, $\xi(r)>$0 for $r<$100 Mpc. Now suppose we only survey a much smaller sub-volume of diameter 50 Mpc within the larger volume. The measured $\xi(r)$ is subject to the constraint that its average must be zero (the total number of pairs is the same - clustering just shifts some pairs to smaller separations). However, we know that the actual $\xi(r)>$0 on all scales in the sub-volume. Hence the measured $\xi(r)$ must be an underestimate and a correction must be made. In other words, because the TONS survey sizes are comparable to scales at which the number density of galaxies in the volume can significantly differ from field to field, we need to account for the fact that they are not representative of the general field. Because the correlation function is only calculated for a finite survey volume, in practise the observed number density $n$ is used instead of the global number density $<n>$. 

Unlike most studies, we have corrected for the sample being under- or over-dense due to large-scale structure. This is possible because, in calculating the model redshift distributions for each subsample, we have already estimated the underlying number density of radio galaxies in the TONS sub-samples. The observed correlation function, $\xi(r)_{\rm obs}$ was modified to the true correlation function, $\xi(r)_{\rm true}$ by a normalisation term, $n/<n>$:

\begin{equation}\label{eq:xi_cor}
(1+\xi(r)_{\rm true})=(1+\xi(r)_{\rm obs})\left(\frac{n}{<n>}\right)^2,
\end{equation}

\noindent \citep{pea} where $n$ is the total number of objects in the sample and $<n>$ is the total number of objects expected in the sample as calculated from our model redshift distribution (see Table.~\ref{tab:the_samples}). In an under-dense region, the number of random-random pairs calculated in the correlation function will be under-estimated and hence the correlation function will be over-estimated and must be corrected downwards. Applying this formula is essentially the same as changing the number of random objects in equation.~\ref{eq:xi} from the number of objects in the sample to the number of objects expected.

%The correlation function may also be affected by edge effects. This effect comes about because the number of data-data pairs in a given separation bin is influenced by the distribution of galaxies relative to the boundaries of the sample. Edge effects are taken into account by including the number of data-random pairs in the calculation of the correlation function (\citealt{ls}; Equation.~\ref{eq:xi}). This effectively measures the average available volume around each data point.

The observed correlation function is also modified because it is measured not in real space, but in redshift space. In practise, the redshift measurements will be affected by large-scale coherent velocities and small-scale virialised motions. This will tend to skew the correlation function and boost the value of $r_0$. The linear analysis of \citet{kai87} implies a boost in the redshift space correlation function by a factor of 1.2 \citep{pea99}. Because of uncertainties in this effect, and because there are other large uncertainties, we have not corrected for this effect.

Finally, to allow for redshift evolution in the clustering of radio galaxies, an evolution term can be incorporated into the definition of the correlation length:

\begin{equation}\label{eq:arr_nort}
r_0(z)=r_0(0)\left(1+z\right)^{-(3+\epsilon-\gamma)/\gamma},
\end{equation}

\noindent (e.g. \citealt{ove}; \citealt{bro03}). $\epsilon$=0 and $\epsilon$=$\gamma$-3 represent scenarios in which the clustering is fixed in physical or co-moving co-ordinates respectively. $\epsilon$=$\gamma$-1 represents growth of clustering under linear perturbation theory \citep{pee}. Unless otherwise stated, in all following analysis, $r_0(0)$ refers to the correlation function at $z$=0 assuming growth of clustering under linear theory. 

\subsection{Calculating the errors on the two-point correlation function}

The Poisson errors on the correlation function at a given separation $r$ are given by:

\begin{equation}
\sigma_\xi(r)=\frac{1+\xi(r)}{\sqrt{DD(r)}}
\end{equation}  

\noindent (e.g. \citealt{pn}). Poisson statistics assume that all pairs are independent of each other which is not the case. The Poisson errors will therefore be an underestimate of the true error. We also used the bootstrap method to estimate the errors (e.g. \citealt{bbs}; \citealt{lbf}; \citealt{mjb}). This involved re-sampling of the data to generate further data sets with a population distribution identical to that of the real data set. We did this by selecting a random sample of $N$ objects from the real catalogue and re-calculating the two-point correlation function for the new sample. The bootstrap estimate of the standard deviation $\sigma_i$ of the re-sampled correlation function $\xi^{*}$ is given by:

\begin{equation}
\sigma_i^2=\sum_{k=0}^{N-1}\frac{(\xi^{*}_{i}-<\xi^{*}_{i}>)^2}{N-1},
\end{equation}

\noindent where $<\xi^{*}>$ is the mean value of $\xi^{*}$ in separation bin $i$. We calculated this for 1000 re-sampled datasets ($N$=1000). Because some galaxies were counted more than once, there were a large number of real-real pairs in the smallest separation bin. Although this is not therefore a good estimate of the mean, it does provide a good estimate of the internal error. 

Because the bootstrap method calculates the internal variance of the sample, the errors will always be an underestimate of the true error as they don't include the error in the underlying number density. However, we have already corrected for this from the model redshift distribution. The only uncertainty will be from the error in our knowledge of the underlying number density. We have already estimated the 1$\sigma$ errors associated with the total number of radio galaxies expected in each sample (Table.~\ref{tab:the_samples}). This was determined from the $\pm$1$\sigma$ errors on the model redshift distribution calculated in Sec.~\ref{sec:nz}. To estimate our total error, we added the fractional errors in quadrature (including twice the fractional error on $<n>$ because $\xi(r)\propto <n>^2$).

\subsection{Estimating $r_0$}
\label{sec:croft}

Because each galaxy can contribute to multiple close pairs, the correlation function obtained in neighbouring bins may be correlated and hence in order to obtain the correlation length it is not strictly correct to simply calculate the best fit to a power-law function (Equation.~\ref{eqn:r0}). Ideally we would obtain the full covariance error matrices using our bootstrap error estimate:

\begin{equation}
\sigma_{ij}^2=\sum_{i=0}^{N-1}\frac{(\xi^{*}_{i}-<\xi^{*}_{i}>)(\xi^{*}_{j}-<\xi^{*}_{j}>)}{N-1}.
\end{equation}

\noindent However, to calculate this, each of the data points must be independent. \citet{mad} overcome this problem by dividing their survey area into contiguous regions and selecting areas at random (instead of the standard bootstrap method in which galaxies are selected at random). A similar approach is that of jackknife re-sampling (see e.g. \citealt{zeh}). Because our sample volume was too small to divide into enough regions (each required to be larger than the separation distances under consideration), we instead chose to estimate the value of the correlation length by using the Levenberg-Marquardt technique to perform a non-linear least squares fit of a power-law function (Equation.~\ref{eqn:r0}) to the binned correlation function data \citep{pre}. In this way, we incorporated the calculated errors rather than assuming simple Poisson errors. Because we were using only the diagonal terms of the covariance matrix, we cannot guarantee that we have obtained an unbiased estimate of the correlation length. However, the goodness of fit suggests that any such bias is negligible. 

\subsection{Relating the correlation length to the bias}

The simplest models relating the galaxy distribution to that of the underlying dark matter is a linear bias: $\rho_{\rm{gal}}= b \times \rho_{\rm{mass}}$ where $b$ is a constant bias factor. The bias is therefore related to the correlation function by:

\begin{equation}
\xi_{\rm{gal}}(r)=b^2 \xi_{\rm{mass}}(r).
\end{equation}

\noindent This can be understood intuitively: clusters of galaxies will form preferentially at the sites of high-density peaks in the primordial density field, and will consequently have a larger bias. 

The bias of radio galaxies $b_{\rm{rg}}$ with respect to the bias of another type of galaxy, $b_{\rm{gal}}$ is therefore related to their scale-lengths $r_0$ and power law indices $\gamma$ by:

\begin{equation}\label{eq:b_r}
\frac {b_{\rm{rg}}}{b_{\rm{gal}}}=\frac{ \left( r_{0 ~\rm{rg}} \right) ^{\gamma_{\rm{rg}}/2} } {\left(  r_{0 ~\rm{gal}}\right) ^{\gamma_{\rm{gal}}/2} }
\end{equation}

\section{Results}

In this section, we discuss the correlation function and its corresponding correlation length obtained for each of the TONS sub-samples. We summarise the correlation lengths obtained along with values obtained in other studies in Table.~\ref{tab:rnort}. 

\begin{table*}
\begin{center}
\begin{tabular}{rrrrrr}
\hline\hline
Survey & mean $z$ & $r_0(z)$ (Mpc $h^{-1}$) & $r_0(z)$ (Mpc) & $r_0$(0) (Mpc) & $b/b_{\rm opt}$ \\
\hline
\citet{nor02} (L$_{\star}$ ellipticals 2dfGRS) & local &5.7$\pm$0.6 & 8.1$\pm$0.9 & {\bf 8.1$\pm$0.9} & 1.0\\
\citet{nor02} (brightest ellipticals 2dfGRS) & $\sim$0.1 &9.7$\pm$1.2 & 13.9$\pm$1.7 & {\bf 15.1$\pm$1.9} & 1.7$\pm$0.3\\
\citet{dal} (APM ARC$>$1 clusters) & local &14.3$\pm$2.4 & 20.4$\pm$3.4 & 20.4$\pm$3.4 & 2.3$\pm$0.5\\
\citet{crof} (APM ARC$>$2 clusters) & local &21.3$\pm^{11.1}_{9.3}$ &  30.4$\pm^{15.9}_{13.3}$ & 30.4$\pm^{15.9}_{13.3}$ & 3.3$\pm^{1.9}_{1.6}$ \\
\citet{pn} (local radio galaxies) & local &11.0$\pm$1.2 & 15.7$\pm$1.7 & 15.7$\pm$1.7 & 1.8$\pm$0.3 \\
\citet{bw} (NVSS radio galaxies) & $\sim$1.0 & -  & -  & 8.6 & 1.05 \\
\citet{mag04} (2dFRS radio galaxies) & $\sim$0.1 & -  & 13.0$\pm$0.9  & 14.2$\pm$1.0 & 1.7$\pm$0.3 \\
TONS08 & $\sim$0.3 & - & 8.0$\pm$2.4 & 10.2$\pm$3.0 & 1.2$\pm$0.4 \\
TONS12 & $\sim$0.3 & - & 11.3$\pm$1.8 & 14.3$\pm$2.3 & 1.7$\pm$0.4 \\
TONS08/12 & $\sim$0.3 & -  & 8.7$\pm$1.6 & {\bf 11.0$\pm$2.0} & 1.3$\pm$0.3 \\
\citet{lac} & $\sim$0.3 &17.9$\pm$7.4 & 12.5$\pm$5.2 & 15.9$\pm$6.6 & 1.5$\pm$0.7 \\
TONS08w/16w/\citet{lac} & $\sim$0.3 & -  & 8.8$\pm$4.4 & 11.2$\pm$5.6 & 1.4$\pm$0.8 \\
\hline\hline
\end{tabular}
{\caption[Surveys]{\label{tab:rnort} Table summarising the measured two-point correlation length, $r_0(z)$ ($h^{-1}$ Mpc), the measured two-point correlation length assuming $h$=0.7, $r_0(z)$ (Mpc), the correlation function corrected to $z$=0 assuming growth of clustering by linear theory, $r_0$(0) (Mpc) and the corresponding bias with respect to L$_{\star}$ elliptical galaxies in the 2dfGRS \citep{nor02}. \citet{nor02} look at clustering properties of different types of galaxies in the 2dfGRS. We take the clustering lengths of L$_{\star}$ ellipticals. We assume this is a local measurement and use this value as  $b_{\rm opt}$. The brightest ellipticals are defined as M$_{\rm b_{j}}$-5log$_{10}h^{-1}$=-21.0-22.0. To convert to $r_0$(0) we assume the median redshift of this population is $z$=0.1. \citet{bw} calculate a rough value for the angular clustering of NVSS ($s_{1.4}>$10 mJy) radio galaxies. The value $r_0(0)$ has already been corrected for growth of clustering under linear theory assuming a median redshift of $z$=1. 
}}
 \end{center}
 \end{table*}

\subsection{The TONS08 and TONS12 sub-samples}

\begin{figure*}
\begin{center}
\setlength{\unitlength}{1mm}
\begin{picture}(150,65)
\put(-2,-10){\includegraphics{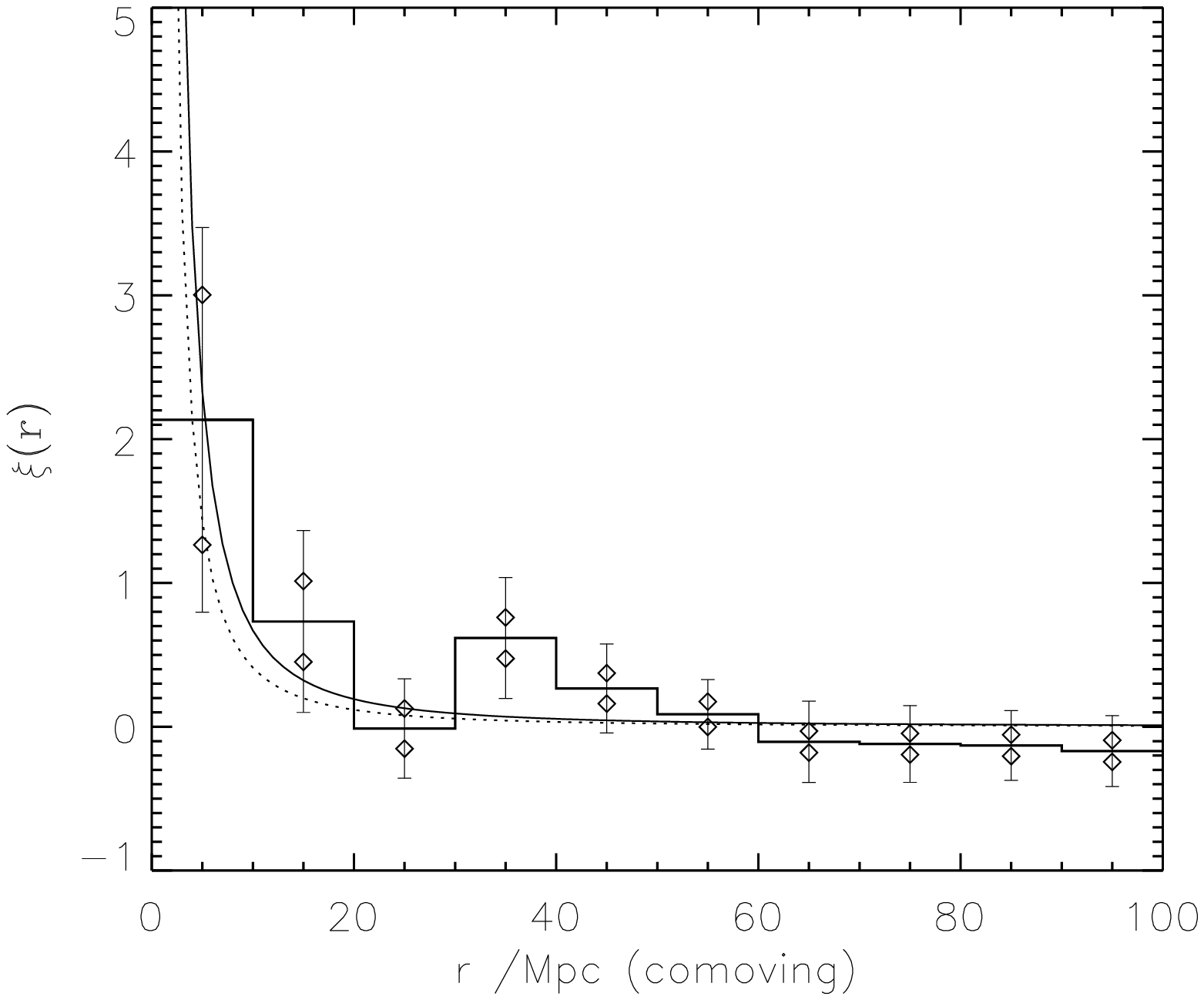}}
\put(70,-10){\includegraphics{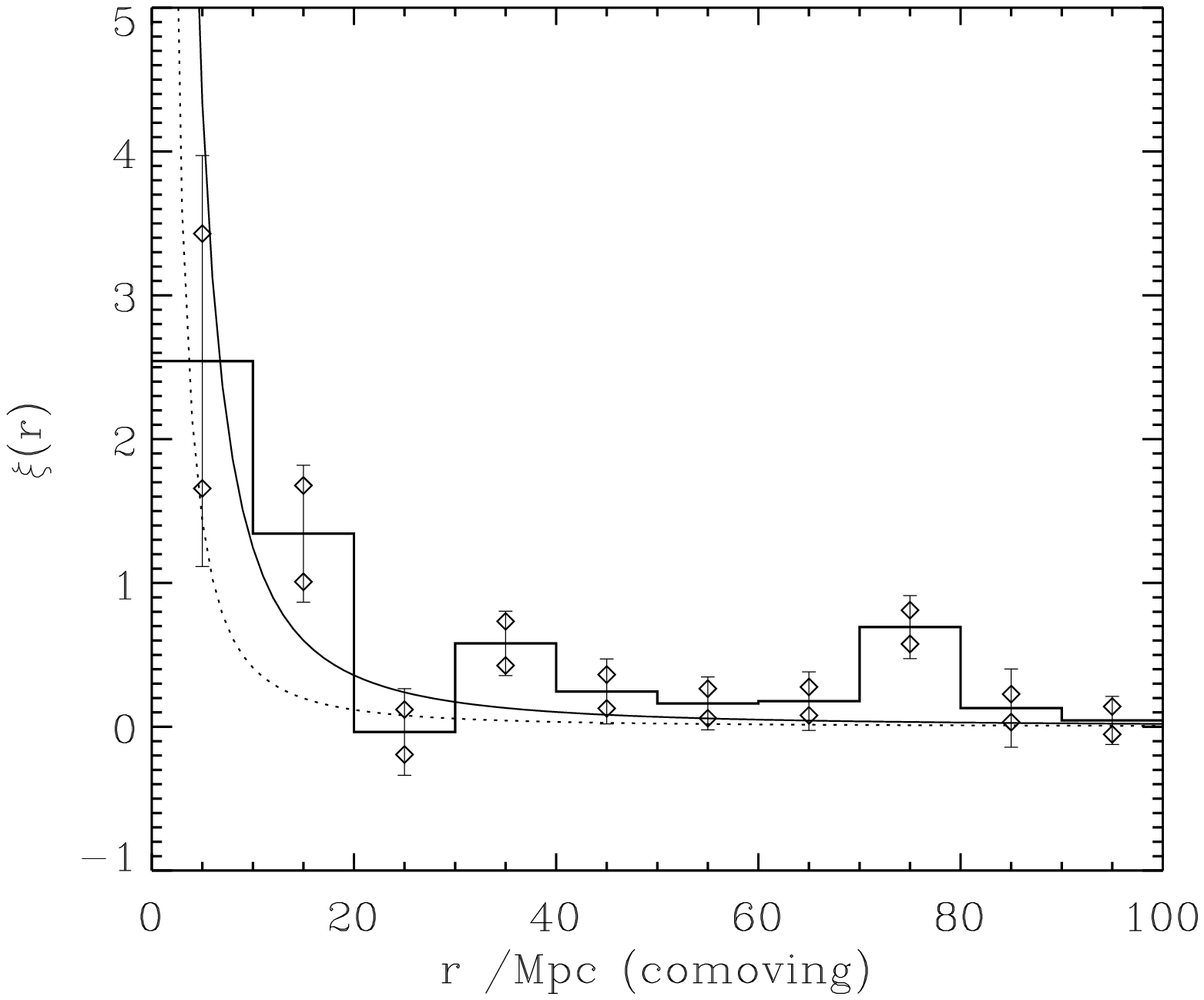}}
\end{picture}
\end{center}
{\caption[junk]{\label{fig:corrfn_08} The two-point correlation function for the TONS08 and TONS12 sub-samples. The data are binned in 10 Mpc separation bins. The errors plotted are obtained from a combination of the bootstrap method and the error due to the uncertainties in the predicted number of radio galaxies. The diamonds show the extent of the Poisson errors. The solid line shows the fitted power-law correlation function with $r_0$=8.0 Mpc (TONS08) and $r_0$=11.3 Mpc (TONS12) and a fixed slope of $\gamma$=1.8. The dotted line shows the correlation function fitted by \citet{pn} for local radio galaxies evolved to $z$=0.3 assuming linear theory ($r_0$=6.1 Mpc).
 }}
\end{figure*}

Fig.~\ref{fig:corrfn_08} shows the two-point correlation function calculated for the TONS08 and TONS12 sub-samples. In the case of TONS08, there are less radio galaxies than predicted. The number of radio galaxies in the random sample is effectively increased and the correlation function has therefore been corrected downwards. Conversely, the TONS12 sample has more radio galaxies than predicted and the correlation function was corrected upwards. 

In both cases, a power-law function provided a poor fit to the correlation function. This is due to the presence of an excess number of pairs of galaxies with separations of 30-80 Mpc and is probably caused by the presence of super-structures in the sub-samples (see Sec.~\ref{sec:ssreg}). The error bars become large in the low separation bins due to small number statistics. The bootstrap errors also become larger when the discrepancy between $n$ and $<n>$ becomes large. This is due to the multiplication factor of $(n/<n>)^2$ in equation.~\ref{eq:xi_cor} for each bootstrap realisation. Because of the large uncertainties, we fixed $\gamma$ to the canonical value of $\gamma$=1.8. We fitted a power-law function to the correlation function out to separations of $r$=200 Mpc. Our best fit values for the correlation length are $r_0(z)$=8.0$\pm$2.4 Mpc and $r_0(z)$=11.3$\pm$1.8 Mpc (1$\sigma$) for the TONS08 and TONS12 sub-samples respectively. The best fit power-law models are overplotted in Fig.~\ref{fig:corrfn_08}. Assuming that the sub-samples have a median redshift of $z$=0.3 and that clustering grows according to linear theory, we used Equation.~\ref{eq:arr_nort} to determine that the equivalent correlation length at $z$=0 are $r_0(0)$=10.2$\pm$3.0 Mpc and $r_0(0)$=14.3$\pm$2.3 Mpc (1$\sigma$). The higher value for the correlation length in the TONS12 sample is probably in part due to the presence of an excess number of pairs of galaxies with separations of 10-20 Mpc which will tend to pull the fitted function up to higher values of $r_0$.

\begin{figure*}
\begin{center}
\setlength{\unitlength}{1mm}
\begin{picture}(150,90)
\put(20,-15){\includegraphics{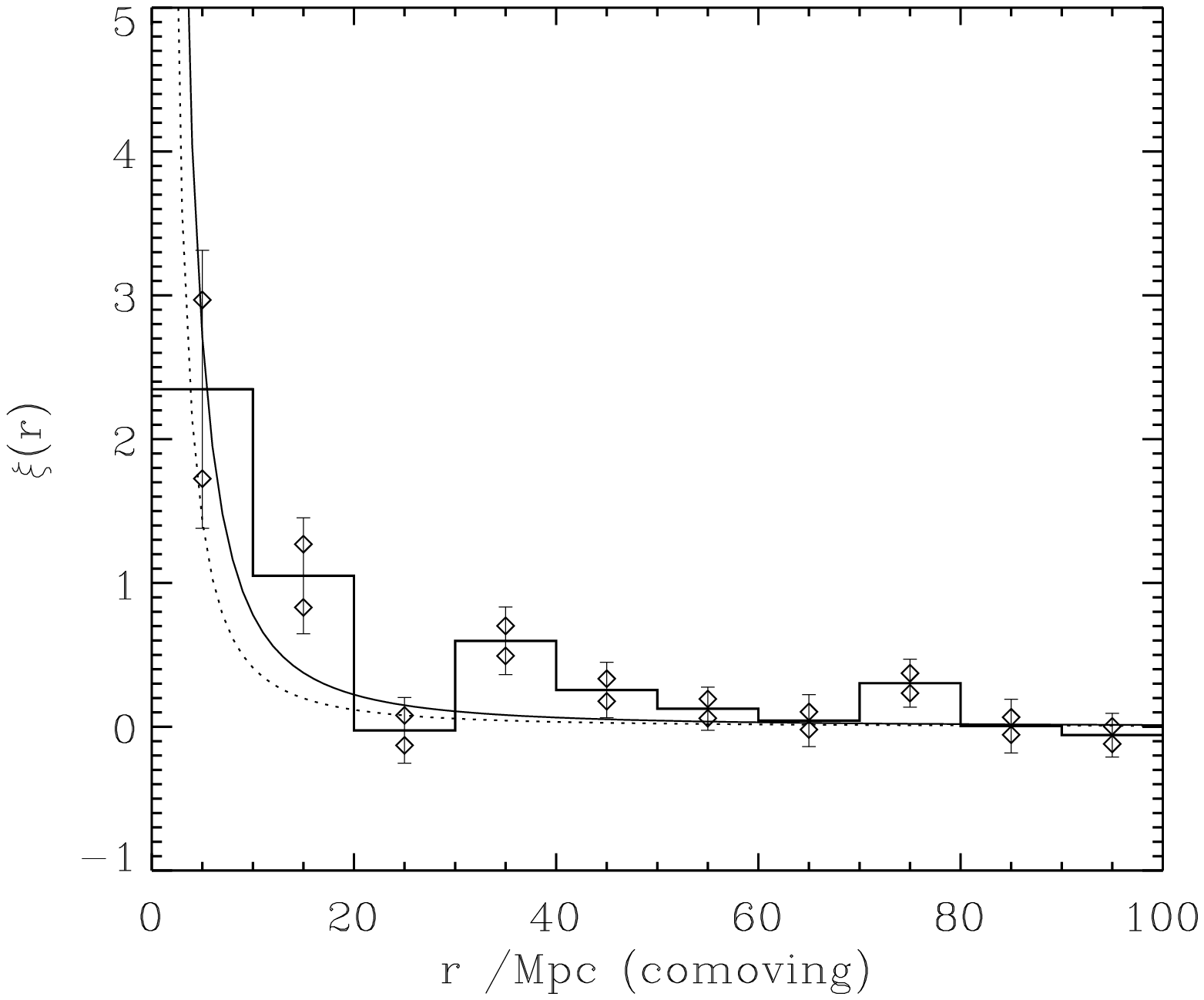}}
\end{picture}
\end{center}
{\caption[junk]{\label{fig:corrfn_08_12} The two-point correlation function for the combined TONS08 and TONS12 sub-samples. The data are binned in 10 Mpc separation bins. The errors plotted are obtained from a combination of the bootstrap method and the error due to the uncertainties in the predicted number of radio galaxies. The diamonds show the extent of the Poisson errors. The solid line shows the fitted power-law correlation function with $r_0$=8.7 Mpc and a fixed slope of $\gamma$=1.8. The dotted line shows the correlation function fitted by \citet{pn} for local radio galaxies evolved to $z$=0.3 assuming linear theory ($r_0$=6.1 Mpc).
 }}
\end{figure*}

To obtain better number statistics, we computed the correlation function for the combined TONS08 and TONS12 sub-samples. Fig.~\ref{fig:corrfn_08_12} shows that the combined correlation function has much smaller errors than is obtained for the separate sub-samples. This is partly because of the increased sample size and also because when TONS08 and TONS12 are combined, the number of radio galaxies observed is very similar to the number expected (see Table.~\ref{tab:the_samples}). 

The best fit correlation length for the combined samples is $r_0(z)$=8.7$\pm$1.6 Mpc (1$\sigma$) (fitting the power-law function out to separations of 30 Mpc). This corresponds to $r_0(0)$=11.0$\pm$2.0 Mpc (1$\sigma$). Using Equation.~\ref{eq:b_r} and taking $r_{0 ~\rm{gal}}(0)$=8.1$\pm$0.9 Mpc \citep{nor02} and $b_{\rm{gal}}$=1.0, this corresponds to a bias of $b_{\rm{rg}}$=1.3$\pm$0.3. 

\subsection{Super-structure regions}
\label{sec:ssreg}

By re-dividing our combined sample into those within and out of super-structure regions, we directly determined if there are any differences in the clustering properties of radio galaxies in different large-scale environments. We combined the TONS08 and TONS12 sub-samples and redivided the combined sample into radio galaxies that were within super-structure regions and those that were not. We defined super-structure members as radio galaxies with redshifts in the intervals $z$=0.233-0.285 and $z$=0.33-0.367 for the TONS08 sub-sample and $z$=0.21-0.25 and $z$=0.285-0.34 for the TONS12 sub-sample. Using this criteria, this resulted in 50/84 and 42/107 radio galaxies defined as being in a super-structure and 34/84 and 65/107 radio galaxies defined as being outside of super-structures in TONS08 and TONS12 respectively. The total sample size was 92 radio galaxies within super-structures and 99 out of super-structures. From our model redshift distribution, we calculated that we would expect 22.3$\pm$2.4 and 24.6$\pm$2.4 radio galaxies within the super-structure regions and 72.8$\pm$3.8 and 73.0$\pm$3.8 radio galaxies outside super-structure regions in the TONS08 and TONS12 sub-samples respectively (see Table.~\ref{tab:the_samples}).   

\begin{figure*}
\begin{center}
\setlength{\unitlength}{1mm}
\begin{picture}(150,65)
\put(-2,-10){\includegraphics{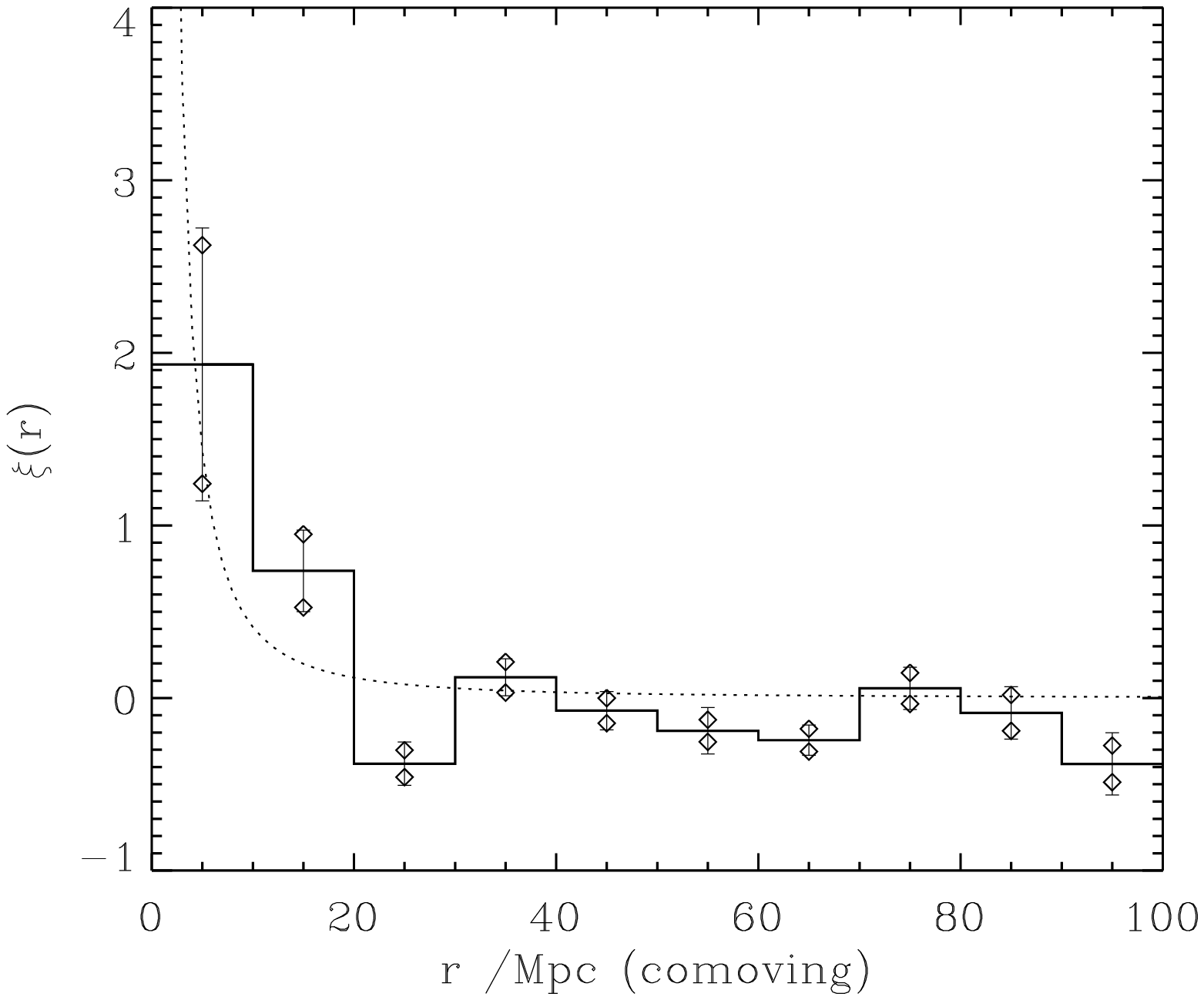}}
\put(70,-10){\includegraphics{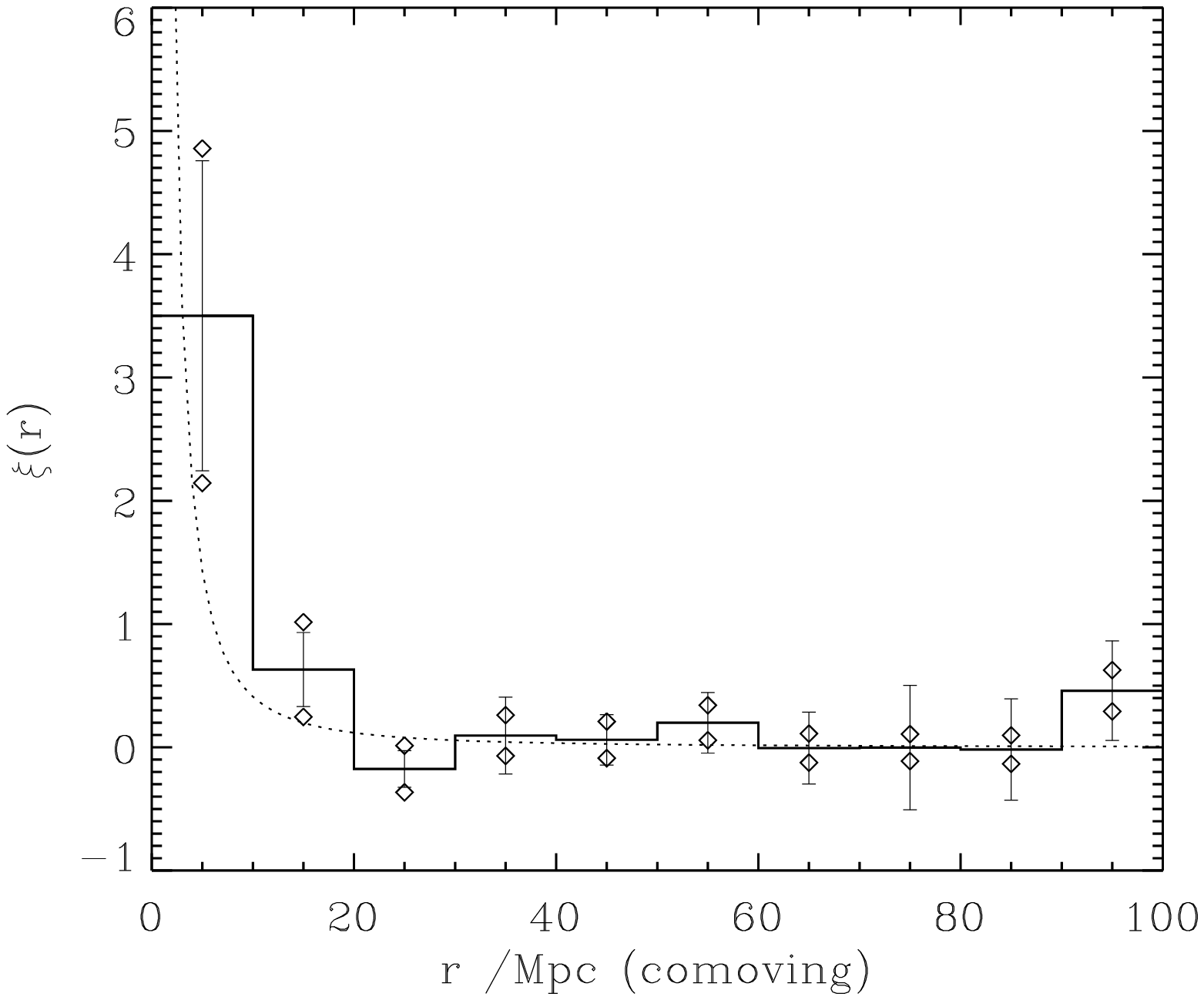}}
\end{picture}
\end{center}
{\caption[junk]{\label{fig:corrfn_08_12_ss} The two-point correlation function for the radio galaxies within (left) and outside of (right) super-structure regions in the combined TONS08 and TONS12 sub-samples. The data are binned in 10 Mpc separation bins. The errors plotted are obtained from a combination of the bootstrap method and the error due to the uncertainties in the predicted number of radio galaxies. The diamonds show the extent of the Poisson errors. The dotted line shows the correlation function fitted by \citet{pn} for local radio galaxies evolved to $z$=0.3 assuming linear theory ($r_0$=6.1 Mpc).
 }}
\end{figure*}

The two-point correlation functions for the combined TONS08 and TONS12 super-structure and non-super-structure regions are shown in Fig.~\ref{fig:corrfn_08_12_ss}. In these cases, we calculated the two-point correlation functions without correcting for the underlying number density of radio galaxies because we were determining differences in the shape and strength of the correlation function in over- and under-dense regions rather than trying to determine a universal correlation length. The two-point correlation functions have very different shapes. The two-point correlation function in non-super-structure regions appears to conform to a well-behaved power-law relation whereas the correlation function in super-structure regions strongly deviates from a simple power-law. The enhanced number of pairs at large separations seen in the correlation functions of both the TONS08 and TONS12 samples are due to radio galaxies in super-structure regions. This is unsurprising given that we would expect a clustering signal on larger scales in regions that we see super-structures. \citet{wild04} use the relative bias of galaxies in the 2dFGRS to rule out an exact linear bias. The deviation from a simple power-law in super-structure regions supports the view that bias may be a more complicated function. 

The two plots illustrate nicely how a power-law is only a good fit either out of super-structure regions or in surveys that are less biased and large enough for their contribution to be negligible. In a survey consisting of biased tracers of the mass, such super-structures will be common and thus a power-law will never be a good approximation. 

\subsection{The TONS08w, TONS16w and Lacy sub-samples}

We also computed the two-point correlation function for the sub-samples with a higher radio flux-density limit. 

\citet{lac} analysed an independent sample of similar ($s_{1.4}>20$ mJy) radio galaxies in a different area of the sky but with a much smaller sample size. To compare our use of a different method and in particular, a different redshift distribution, we compared the correlation length that we obtain with that obtained by \citet{lac}. Without correcting for the fact that the region is under-dense, we obtained a correlation length of $r_0(z)=15.3\pm7.1$ Mpc. The correlation length obtained by \citet{lac} is $r_0=17^{+7}_{-12}$ Mpc. Our results agree with this well within the error bars. When we corrected the correlation function to that for the underlying number density, we found $r_0(z)=12.5\pm5.2$ Mpc. This shows that the effects of calculating $\xi$ over a small survey volume are large. Assuming a median redshift of $z$=0.3, this corresponds to a correlation length at $z$=0 of $r_0(0)=15.9\pm6.6$ Mpc. 

As demonstrated for the \citet{lac} sample, the small numbers combined with the smaller number density of the higher flux density radio galaxies in the TONS08w and TONS16w sub-samples means that the errors are too large to determine a reliable value of $r_0$ for each individual sub-sample. We therefore combined the TONS08w, TONS16w and \citet{lac} samples to improve the small number statistics. Although the \citet{lac} sample has slightly different selection criterion, the populations are sufficiently similar to add the samples together to increase the number statistics and still obtain meaningful results. 

\begin{figure}
\begin{center}
\setlength{\unitlength}{1mm}
\begin{picture}(150,65)
\put(-2,-10){\includegraphics{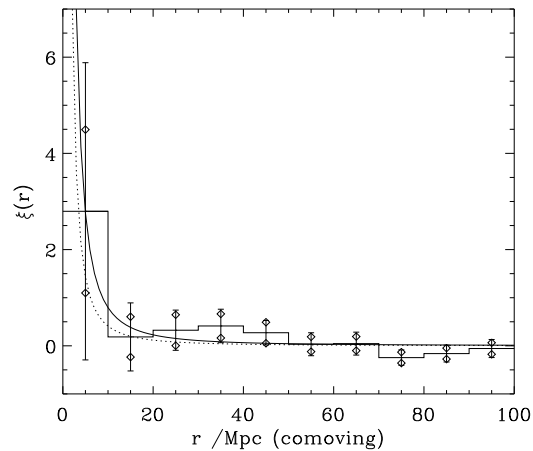}}
\end{picture}
\end{center}
{\caption[junk]{\label{fig:corrfn_highf} The two-point correlation function for the combined TONS08w, TONS16w and \citet{lac} samples assuming that the predicted number of radio galaxies in TONS08w is equal to the number observed. The data are binned in 10 Mpc separation bins. The errors plotted are obtained from a combination of the bootstrap method and the error due to the uncertainties in the predicted number of radio galaxies. The diamonds show the extent of the Poisson errors. The solid line shows the power-law correlation function with $r_0$=8.8 Mpc and a fixed slope of $\gamma$=1.8. The dotted line shows the correlation function fitted by \citet{pn} for local radio galaxies evolved to $z$=0.3 assuming linear theory ($r_0$=6.1 Mpc).
 }}
\end{figure}

The combined two-point correlation function is shown in Fig.~\ref{fig:corrfn_highf}. Because of the possibility of incompletenesses in the TONS08w sub-sample, we assumed that the underlying number density of radio galaxies in the TONS08w survey was equal to the number observed (Equation.~\ref{eq:xi_cor}). The error bars are too large to obtain a reasonable estimate of $r_0$. A number of reasons may be accountable for the lack of signal and the large error bars. An intrinsic reason may be the smaller number density of the higher flux density radio galaxies. Perhaps there are too few luminous radio galaxies triggered at small separations. Indeed, there are only four, one and zero radio galaxy pairs with separations $<$ 10 Mpc in the \citet{lac}, TONS16w and TONS08w sub-samples respectively. The \citet{lac} sample gives a particularly high clustering signal (even after it is corrected for being an under-dense region). This is in contrast to the TONS08w and TONS16w sub-samples which appear to have much weaker clustering signals. 

We obtained a clustering length of $r_0(z)$=8.8$\pm$4.4 Mpc (1$\sigma$). Assuming a median redshift of $z$=0.3, this corresponds to a correlation length at $z$=0 of $r_0(0)$=11.2$\pm$5.6 Mpc (1$\sigma$). This is consistent with the findings of \citet{mag04} who found no difference in the clustering properties of radio-AGNs of different radio luminosity in the 2dFGRS and \citet{ove} who found that the angular correlation function of NVSS radio galaxies is approximately constant between flux density limits of $s_{1.4}$=3 - 40 mJy. \citet{mag04} concluded that once radio activity is triggered in low luminosity radio galaxies, there is no evidence for a connection between black-hole mass and the level of radio output. However our errors are so large that one can rule out only very dramatic differences in the clustering properties of the two samples. 

\section{DISCUSSION}
\label{sec:discussion}

\begin{figure*}
\begin{center}
\setlength{\unitlength}{1mm}
\begin{picture}(150,90)
\put(20,-15){\includegraphics{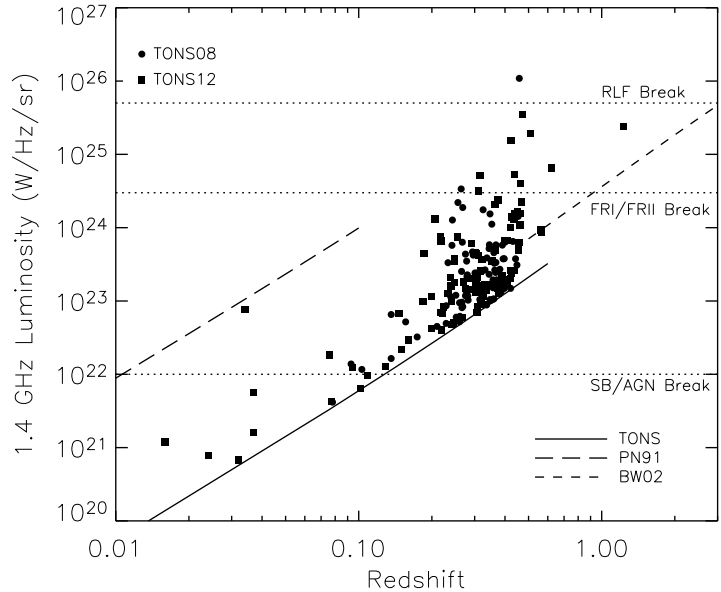}}
\end{picture}
\end{center}
{\caption[junk]{\label{fig:lz_relation} 1.4 GHz Luminosity as a function of redshift for the TONS08 (circles) and TONS12 (squares) sub-samples. Also plotted is the luminosity corresponding to the radio galaxy flux density limits over the appropriate redshift range for the TONS08/12 sub-samples ($s_{\rm 1.4}$=3 mJy; solid line), the \citet{pn} sample ($s_{\rm 1.4}$=0.5 Jy; long dashed line) and the \citet{bw} sample ($s_{\rm 1.4}$=10 mJy; short dashed line). Various critical values of the radio luminosity are marked: the star-burst (SB) / AGN break ($L_{\rm 1.4 GHz} \sim 10^{22} {\rm W Hz^{-1} sr^{-1}}$); the FRI/FRII break ($L_{\rm 1.4 GHz} \sim 3 \times 10^{24} {\rm W Hz^{-1} sr^{-1}}$) and the RLF break ($L_{\rm 1.4 GHz} \sim 10^{25.7} {\rm W Hz^{-1} sr^{-1}}$).
 }}
\end{figure*}

By combining the TONS08 and TONS12 sub-samples, we found a correlation length of $r_0(z)$=8.7$\pm$1.6 Mpc. Even after correcting for the increase in clustering under linear theory between $z\sim$0.3 and the local Universe ($r_0(0)$=11.0$\pm$2.0 Mpc), this value is significantly smaller than the local value of $r_0(0)$=15.7$\pm$1.7 Mpc derived by \citep{pn} and that obtained by \citet{mag04} ($r_0(0)$=14.2$\pm$1.0 Mpc). The reason for this discrepancy is unclear as we should be looking at a comparable sample in terms of the mean radio luminosity of the population (see Fig.~\ref{fig:lz_relation}). This may be evidence for a redshift evolution in the clustering of low luminosity radio galaxies but we note that the sign of the supposed effect (radio galaxies are less biased at $z \sim$0.3 than at $z \sim$0) runs counter to the inferences drawn from systematic changes in the environment of weak radio galaxies with redshift \citep{hl}. However, a number of different effects could contribute to this apparent discrepancy including a marginal statistical difference (see errors on biases in Table.~\ref{tab:rnort}), small number statistics, redshift space distortion effects and the assumptions that have to be adopted when transforming the clustering measurements at different redshifts to that in the local Universe. 

We obtained a larger correlation length than that obtained from the projection of the angular correlation function of NVSS radio galaxies (\citealt{bw}; \citealt{ove}). Any discrepancy in this case is easier to explain: Fig.~\ref{fig:lz_relation} illustrates how the \citet{bw} sample is dominated by more powerful FRII radio galaxies near the break in the RLF. \citet{mclu} showed that although the scale size and optical luminosity of the host galaxy is correlated with the radio power for FRII's, this relation does not hold for FRI radio galaxies. If FRI radio galaxies have more luminous host galaxies, they should also have a larger clustering length. The radio activity in these populations may be triggered by a different mechanism. This is hinted at by the recent result of \citet{cj} who found that although it is well known that the co-moving number density of powerful FRII radio galaxies falls dramatically at low redshifts, the co-moving number density of lower luminosity AGN remains constant. The clustering strength of FRII radio galaxies may therefore be more complex than simply tracing the clustering of the underlying host galaxy population. However, these measurements may have significantly underestimated the correlation length for two reasons. Firstly, deep cleaning of the NVSS images may have caused a deficit of galaxy pairs with separations $\theta\sim$0.2 deg, which may have resulted in a spuriously low power-law fit and hence correlation length \citep{bla04}. Secondly, the assumed redshift distribution used to deproject the angular correlation function has an unphysical spike at low redshifts which would also have suppressed the correlation signal. When these effects are accounted for, the correlation length obtained was $r_0(0)$=15.1$\pm$0.7 (C. Blake, private communication).

\citet{nor01} showed that the strength of galaxy clustering depends on the galaxies intrinsic optical luminosity. The mean luminosity of the TONS08/12 radio galaxies is $\sim$2.5 $L_\star$. From the relation: $b/b_\star=0.85+0.15L/L_\star$ \citet{nor01}, we expect the sample to have a relative bias of $b$=1.2. From \citet{nor02} fig.10, the expected bias of 2.5$L_\star$ early-type galaxies has a slightly higher bias of $b$=1.3. Using equation.10, the TONS08/12 (linear theory evolution corrected) clustering length corresponds to a bias of $b$=1.3$\pm$0.3. Although the uncertainties are large, this suggests that the clustering strength of the TONS radio galaxies is consistent with that of the underlying host galaxy population. If, as suggested by the results of \citet{mclu}, the clustering signal measured by \citet{bw} and \citet{ove} was dominated by galaxies with smaller ($\sim L_\star$) optical luminosities, we would indeed expect them to measure a smaller clustering signal. 

\citet{brand} postulated that the radio galaxy bias of low luminosity radio galaxies needs to be $b\sim$3-4 to explain the number of super-structures found within the TONS08 volume. In this paper, we have shown that redshift spikes are a common phenomena in radio galaxies redshift surveys at NVSS flux density levels and $z\sim$0.3 (corresponding to FRI-like radio galaxies well below the RLF break). The discrepancy in the bias inferred here and by \citet{brand} may be related to the inadequacy of a simple linear bias model for the way radio galaxies trace the underlying dark matter.

\section{CONCLUSIONS}

The results of this investigation can be summarized as follows.

\begin{itemize}
\item We have demonstrated that redshift spikes corresponding to super-structure size overdensities are a common phenomenon within radio galaxy redshift surveys. 
\item We computed a universal redshift distribution for radio galaxy redshift surveys of given optical magnitude and radio flux density limits. 
\item The spatial clustering of low luminosity (predominantly FRI) radio galaxies at $z\sim$0-0.5 has been measured directly for the first time. 
\item We used our model redshift distribution to correct the clustering signal for our sample being under- or over-dense due to large-scale structure.
\item The best fit correlation length for our low flux density limit ($s_{1.4}>$3 mJy) sample is $r_0(z)$=8.7$\pm$1.6 Mpc, corresponding to $r_0(0)$=11.0$\pm$2.0 Mpc at $z$=0 (assuming an increase in clustering under linear theory). This is consistent with the clustering strength of the underlying host galaxy population.
\item  We found that the clustering of low luminosity radio galaxies is not significantly different for our low and high flux density limited samples.
\item By comparing the two-point correlation function for radio galaxies within and outside of super-structure regions, we showed that a power-law function is only a good approximation outside super-structure regions or in surveys that are less biased and large enough for their contribution to be negligible. In a survey consisting of biased tracers of the mass, such super-structures will be common and thus a power-law will never be a good approximation.
\end{itemize}

\section{ACKNOWLEDGEMENTS}

We thank Chris Blake, Michael Brown, Idit Zehavi, Mark Lacy and John Peacock for useful discussions and Devinder Sivia, Will Percival, Jenny Grimes and Steve Croft for use of code and software. We also thank Amanda Bauer, Filipe Abdalla and Ewan Mitchell for obtaining some of the spectra and Chris Benn and Jasper Wall for redshifts of unpublished spectra of some of the radio galaxies in TONS12. 
We thank the staff at McDonald, NOT and the WHT for the  technical support they have provided. This research has made use of APM, NVSS, FIRST \& POSS-II web-based databases. We thank the staff of the Hobby-Eberly Telescope. The HET is operated by McDonald Observatory on behalf of The University of Texas at Austin, the Pennsylvania State University, Stanford University, Ludwig-Maximilians-Universitaet Muenchen, and Georg-August-Universitaet Goettingen.
The Nordic Optical Telescope is operated on the island of La Palma jointly by Denmark, Finland, Iceland, Norway, and Sweden, in the Spanish Observatorio del Roque de los Muchachos of the Instituto de Astrofisica de Canarias.
The WHT is operated on the island of La Palma by the Isaac Newton Group in the Spanish Observatorio del Roque de los Muchachos of the Instituto de Astrofisica de Canarias. This material is based in part upon work supported by the Texas Advanced Research Program under Grant No. 009658-0710-1999.
Kate Brand was supported for the majority of this work by a PPARC studentship. Steve Rawlings thanks PPARC for a Senior Research Fellowship.

\bibliographystyle{mn2e}
\bibliography{thesis_references}

\begin{thebibliography}{48}
\expandafter\ifx\csname natexlab\endcsname\relax\def\natexlab#1{#1}\fi

\bibitem[{{Abazajian} {et~al.}(2004){Abazajian}, {Adelman-McCarthy}, {Ag{\"
  u}eros}, {Allam}, {Anderson}, {Anderson}, {Annis}, {Bahcall}, \& {the SDSS
  collaboration}}]{aba04}
{Abazajian} K., {Adelman-McCarthy} J.~K., {Ag{\" u}eros} M.~A., {Allam} S.~S.,
  {Anderson} K.~S.~J., {Anderson} S.~F., {Annis} J., {Bahcall} N.~A., {the SDSS
  collaboration}, 2004, \aj, 128, 502

\bibitem[{{Barrow} {et~al.}(1984){Barrow}, {Bhavsar}, \& {Sonoda}}]{bbs}
{Barrow} J.~D., {Bhavsar} S.~P., {Sonoda} D.~H., 1984, \mnras, 210, 19P

\bibitem[{{Becker} {et~al.}(1995){Becker}, {White}, \& {Helfand}}]{bwh}
{Becker} R.~H., {White} R.~L., {Helfand} D.~J., 1995, \apj, 450, 559

\bibitem[{{Blake} {et~al.}(2004){Blake}, {Mauch}, \& {Sadler}}]{bla04}
{Blake} C., {Mauch} T., {Sadler} E.~M., 2004, \mnras, 347, 787

\bibitem[{{Blake} \& {Wall}(2002)}]{bw}
{Blake} C., {Wall} J., 2002, \mnras, 329, L37

\bibitem[{{Brand} {et~al.}(2003){Brand}, {Rawlings}, {Hill}, {Lacy},
  {Mitchell}, \& {Tufts}}]{brand}
{Brand} K., {Rawlings} S., {Hill} G.~J., {Lacy} M., {Mitchell} E., {Tufts} J.,
  2003, \mnras, 344, 283

\bibitem[{{Brown} {et~al.}(2003){Brown}, {Dey}, {Jannuzi}, {Lauer}, {Tiede}, \&
  {Mikles}}]{bro03}
{Brown} M.~J.~I., {Dey} A., {Jannuzi} B.~T., {Lauer} T.~R., {Tiede} G.~P.,
  {Mikles} V.~J., 2003, \apj, 597, 225

\bibitem[{{Bruzual} \& {Charlot}(1993)}]{bc}
{Bruzual} A.~G., {Charlot} S., 1993, \apj, 405, 538

\bibitem[{{Clewley} \& {Jarvis}(2004)}]{cj}
{Clewley} L., {Jarvis} M.~J., 2004, \mnras, 352, 909

\bibitem[{{Colless} {et~al.}(2001){Colless}, {Dalton}, {Maddox}, {Sutherland},
  {Norberg}, {Cole}, {Bland-Hawthorn}, {Bridges}, \& {the 2dF
  collaboration}}]{col}
{Colless} M., {Dalton} G., {Maddox} S., {Sutherland} W., {Norberg} P., {Cole}
  S., {Bland-Hawthorn} J., {Bridges} T., {the 2dF collaboration}, 2001, \mnras,
  328, 1039

\bibitem[{{Condon} {et~al.}(1998){Condon}, {Cotton}, {Greisen}, {Yin},
  {Perley}, {Taylor}, \& {Broderick}}]{con}
{Condon} J.~J., {Cotton} W.~D., {Greisen} E.~W., {Yin} Q.~F., {Perley} R.~A.,
  {Taylor} G.~B., {Broderick} J.~J., 1998, \aj, 115, 1693

\bibitem[{{Croft} {et~al.}(1997){Croft}, {Dalton}, {Efstathiou}, {Sutherland},
  \& {Maddox}}]{crof}
{Croft} R.~A.~C., {Dalton} G.~B., {Efstathiou} G., {Sutherland} W.~J., {Maddox}
  S.~J., 1997, \mnras, 291, 305

\bibitem[{{Dalton} {et~al.}(1994){Dalton}, {Croft}, {Efstathiou}, {Sutherland},
  {Maddox}, \& {Davis}}]{dal}
{Dalton} G.~B., {Croft} R.~A.~C., {Efstathiou} G., {Sutherland} W.~J., {Maddox}
  S.~J., {Davis} M., 1994, \mnras, 271, L47+

\bibitem[{{Dunlop} \& {Peacock}(1990)}]{dp}
{Dunlop} J.~S., {Peacock} J.~A., 1990, \mnras, 247, 19

\bibitem[{{Fisher} {et~al.}(1994){Fisher}, {Davis}, {Strauss}, {Yahil}, \&
  {Huchra}}]{fis}
{Fisher} K.~B., {Davis} M., {Strauss} M.~A., {Yahil} A., {Huchra} J., 1994,
  \mnras, 266, 50

\bibitem[{{Folkes} {et~al.}(1999){Folkes}, {Ronen}, {Price}, {Lahav},
  {Colless}, {Maddox}, {Deeley}, {Glazebrook}, \& {the 2dF
  collaboration}}]{fol}
{Folkes} S., {Ronen} S., {Price} I., {Lahav} O., {Colless} M., {Maddox} S.,
  {Deeley} K., {Glazebrook} K., {the 2dF collaboration}, 1999, \mnras, 308, 459

\bibitem[{{Frei} \& {Gunn}(1994)}]{fre}
{Frei} Z., {Gunn} J.~E., 1994, \aj, 108, 1476

\bibitem[{{Groth} \& {Peebles}(1977)}]{gp}
{Groth} E.~J., {Peebles} P.~J.~E., 1977, \apj, 217, 385

\bibitem[{{Hill} \& {Lilly}(1991)}]{hl}
{Hill} G.~J., {Lilly} S.~J., 1991, \apj, 367, 1

\bibitem[{{Hill} {et~al.}(1998){Hill}, {Nicklas}, {MacQueen}, {Tejada de},
  {Cobos}, \& {Mitsch}}]{hill98}
{Hill} G.~J., {Nicklas} H., {MacQueen} P.~J., {Tejada de} V.~C., {Cobos} D.,
  {Mitsch} W., 1998, Optical Instrumentation, S. D'Odorico, Ed., Proc. SPIE
  3355, 375

\bibitem[{{Hill} \& {Rawlings}(2003)}]{hr}
{Hill} G.~J., {Rawlings} S., 2003, New Astronomy Review, 47, 373

\bibitem[{{Hill} {et~al.}(2002){Hill}, {Wolf}, {Tufts}, \& {Smith}}]{hill02}
{Hill} G.~J., {Wolf} M.~J., {Tufts} J.~R., {Smith} E.~C., 2002, Specialized
  Optical Developments in Astronomy, E. Atad-Ettedgui and S. D'Odorico, Eds.,
  Proc. SPIE 4842, 1

\bibitem[{{Kaiser}(1984)}]{kai}
{Kaiser} N., 1984, \apjl, 284, L9

\bibitem[{{Kaiser}(1987)}]{kai87}
---, 1987, \mnras, 227, 1

\bibitem[{{Kerscher} {et~al.}(2000){Kerscher}, {Szapudi}, \& {Szalay}}]{ker}
{Kerscher} M., {Szapudi} I., {Szalay} A.~S., 2000, \apjl, 535, L13

\bibitem[{{Lacy}(2000)}]{lac}
{Lacy} M., 2000, \apjl, 536, L1

\bibitem[{{Landy} \& {Szalay}(1993)}]{lasz}
{Landy} S.~D., {Szalay} A.~S., 1993, \apj, 412, 64

\bibitem[{{Ling} {et~al.}(1986){Ling}, {Barrow}, \& {Frenk}}]{lbf}
{Ling} E.~N., {Barrow} J.~D., {Frenk} C.~S., 1986, \mnras, 223, 21P

\bibitem[{{Madgwick} {et~al.}(2003){Madgwick}, {Hawkins}, {Lahav}, {Maddox},
  {Norberg}, {Peacock}, {Baldry}, {Baugh}, \& {the 2dFGRS collaboration}}]{mad}
{Madgwick} D.~S., {Hawkins} E., {Lahav} O., {Maddox} S., {Norberg} P.,
  {Peacock} J.~A., {Baldry} I.~K., {Baugh} C.~M., {the 2dFGRS collaboration},
  2003, \mnras, 344, 847

\bibitem[{{Magliocchetti} {et~al.}(2004){Magliocchetti}, {Maddox}, {Hawkins},
  {Peacock}, {Bland-Hawthorn}, {Bridges}, {Cannon}, {Cole}, \& {the 2dF
  collaboration}}]{mag04}
{Magliocchetti} M., {Maddox} S.~J., {Hawkins} E., {Peacock} J.~A.,
  {Bland-Hawthorn} J., {Bridges} T., {Cannon} R., {Cole} S., {the 2dF
  collaboration}, 2004, \mnras, 350, 1485

\bibitem[{{Marshall} {et~al.}(1983){Marshall}, {Tananbaum}, {Huchra},
  {Zamorani}, {Braccesi}, \& {Zitelli}}]{mar}
{Marshall} H.~L., {Tananbaum} H., {Huchra} J.~P., {Zamorani} G., {Braccesi} A.,
  {Zitelli} V., 1983, \apj, 269, 42

\bibitem[{{McLure} {et~al.}(2004){McLure}, {Willott}, {Jarvis}, {Rawlings},
  {Hill}, {Mitchell}, {Dunlop}, \& {Wold}}]{mclu}
{McLure} R.~J., {Willott} C.~J., {Jarvis} M.~J., {Rawlings} S., {Hill} G.~J.,
  {Mitchell} E., {Dunlop} J.~S., {Wold} M., 2004, \mnras, 351, 347

\bibitem[{{McMahon} {et~al.}(2002){McMahon}, {White}, {Helfand}, \&
  {Becker}}]{mcm}
{McMahon} R.~G., {White} R.~L., {Helfand} D.~J., {Becker} R.~H., 2002, \apjs,
  143, 1

\bibitem[{{Mo} {et~al.}(1992){Mo}, {Jing}, \& {Boerner}}]{mjb}
{Mo} H.~J., {Jing} Y.~P., {Boerner} G., 1992, \apj, 392, 452

\bibitem[{{Norberg} {et~al.}(2002){Norberg}, {Baugh}, {Hawkins}, {Maddox},
  {Madgwick}, {Lahav}, {Cole}, {Frenk}, \& {the 2dFGRS collaboration}}]{nor02}
{Norberg} P., {Baugh} C.~M., {Hawkins} E., {Maddox} S., {Madgwick} D., {Lahav}
  O., {Cole} S., {Frenk} C.~S., {the 2dFGRS collaboration}, 2002, \mnras, 332,
  827

\bibitem[{{Norberg} {et~al.}(2001){Norberg}, {Baugh}, {Hawkins}, {Maddox},
  {Peacock}, {Cole}, {Frenk}, {Bland-Hawthorn}, \& {the 2dFGRS
  collaboration}}]{nor01}
{Norberg} P., {Baugh} C.~M., {Hawkins} E., {Maddox} S., {Peacock} J.~A., {Cole}
  S., {Frenk} C.~S., {Bland-Hawthorn} J., {the 2dFGRS collaboration}, 2001,
  \mnras, 328, 64

\bibitem[{{Overzier} {et~al.}(2003){Overzier}, {R{\" o}ttgering}, {Rengelink},
  \& {Wilman}}]{ove}
{Overzier} R.~A., {R{\" o}ttgering} H.~J.~A., {Rengelink} R.~B., {Wilman}
  R.~J., 2003, \aap, 405, 53

\bibitem[{{Peacock}(1999{\natexlab{a}})}]{pea99}
{Peacock} J., 1999{\natexlab{a}}, To appear in proceedings of the KNAW
  colloquium - The most distant radio galaxies, Amsterdam, October 1997.

\bibitem[{{Peacock}(1999{\natexlab{b}})}]{pea}
{Peacock} J.~A., 1999{\natexlab{b}}, {Cosmological physics}. Cosmological
  physics.~ Publisher: Cambridge, UK: Cambridge University Press, 1999.~ISBN:
  0521422701

\bibitem[{{Peacock} \& {Nicholson}(1991)}]{pn}
{Peacock} J.~A., {Nicholson} D., 1991, \mnras, 253, 307

\bibitem[{{Peebles}(1980)}]{pee}
{Peebles} P.~J.~E., 1980, {The large-scale structure of the universe}. Research
  supported by the National Science Foundation.~Princeton, N.J., Princeton
  University Press, 1980.~435 p.

\bibitem[{{Press} {et~al.}(1992){Press}, {Teukolsky}, {Vetterling}, \&
  {Flannery}}]{pre}
{Press} W.~H., {Teukolsky} S.~A., {Vetterling} W.~T., {Flannery} B.~P., 1992,
  {Numerical recipes in C. The art of scientific computing}. Cambridge:
  University Press, |c1992, 2nd ed.

\bibitem[{{Sadler} {et~al.}(2002){Sadler}, {Jackson}, {Cannon}, {McIntyre},
  {Murphy}, {Bland-Hawthorn}, {Bridges}, {Cole}, \& {the 2dF
  collaboration}}]{sad}
{Sadler} E.~M., {Jackson} C.~A., {Cannon} R.~D., {McIntyre} V.~J., {Murphy} T.,
  {Bland-Hawthorn} J., {Bridges} T., {Cole} S., {the 2dF collaboration}, 2002,
  \mnras, 329, 227

\bibitem[{{Steidel} {et~al.}(1998){Steidel}, {Adelberger}, {Dickinson},
  {Giavalisco}, {Pettini}, \& {Kellogg}}]{ste}
{Steidel} C.~C., {Adelberger} K.~L., {Dickinson} M., {Giavalisco} M., {Pettini}
  M., {Kellogg} M., 1998, \apj, 492, 428

\bibitem[{{Wild} {et~al.}(2004){Wild}, {Peacock}, {Lahav}, {Conway}, {Maddox},
  {Baldry}, {Baugh}, \& {the 2dFGRS collaboration}}]{wild04}
{Wild} V., {Peacock} J.~A., {Lahav} O., {Conway} E., {Maddox} S., {Baldry}
  I.~K., {Baugh} C.~M., {the 2dFGRS collaboration}, 2004, \mnras, submitted

\bibitem[{{Willott} {et~al.}(2001){Willott}, {Rawlings}, {Blundell}, {Lacy}, \&
  {Eales}}]{wil01}
{Willott} C.~J., {Rawlings} S., {Blundell} K.~M., {Lacy} M., {Eales} S.~A.,
  2001, \mnras, 322, 536

\bibitem[{{Willott} {et~al.}(2002){Willott}, {Rawlings}, {Blundell}, {Lacy},
  {Hill}, \& {Scott}}]{wil02}
{Willott} C.~J., {Rawlings} S., {Blundell} K.~M., {Lacy} M., {Hill} G.~J.,
  {Scott} S.~E., 2002, \mnras, 335, 1120

\bibitem[{{Zehavi} {et~al.}(2002){Zehavi}, {Blanton}, {Frieman}, {Weinberg},
  {Mo}, {Strauss}, {Anderson}, {Annis}, \& {the SLOAN collaboration}}]{zeh}
{Zehavi} I., {Blanton} M.~R., {Frieman} J.~A., {Weinberg} D.~H., {Mo} H.~J.,
  {Strauss} M.~A., {Anderson} S.~F., {Annis} J., {the SLOAN collaboration},
  2002, \apj, 571, 172

\end{thebibliography}

\end{document}